\documentclass[11pt]{article}

\usepackage{graphicx,epstopdf,float,multirow,booktabs}
\usepackage{amsmath,amssymb,amsfonts,amsthm}
\usepackage{mathrsfs}
\usepackage{mathtools}
\usepackage{array}
\usepackage{xcolor}
\usepackage{textcomp}
\usepackage{url}
\usepackage{comment}
\usepackage{enumitem}
\usepackage{subcaption}
\usepackage{caption}
\usepackage[title]{appendix}
\usepackage{tikz}
\usetikzlibrary{positioning, calc, arrows.meta, patterns}
\usepackage{pgfplots}
\pgfplotsset{compat=1.18}
\usepgfplotslibrary{groupplots}
\usepackage[letterpaper,margin=1in]{geometry}
\usepackage{authblk}
\usepackage[authoryear]{natbib}
\usepackage{hyperref}
\usepackage{cleveref}

\def\Pr{\mathrm{Pr}}

\DeclareMathOperator{\logit}{logit}
\newcommand*\diff{\mathop{}\!\mathrm{d}}

\theoremstyle{definition}
\newtheorem{prop}{Proposition}

\newtheorem{corollary}{Corollary}

\newtheorem{definition}{Definition}
\theoremstyle{plain}

\theoremstyle{remark}

\begin{document}

\title{A Tale of Two Pathways to Gompertz Mortality:\\ Reliability and Vitality from an Actuarial Perspective}

\author[1]{Guoqian Li}
\author[1]{Kenneth Q. Zhou\thanks{Corresponding author. E-mail: kenneth.zhou@uwaterloo.ca}}
\author[2]{Xiaobai Zhu}
\affil[1]{University of Waterloo, Waterloo, Canada}
\affil[2]{The Chinese University of Hong Kong, Hong Kong, China}
\date{\today}

\maketitle

\begin{abstract}
This paper studies two mechanistic explanations for human mortality by examining reliability theory and vitality modelling through a unified actuarial perspective. While the two approaches arise from different ageing mechanisms, we show that both can naturally generate the Gompertz law under suitable assumptions and can be extended to produce the Makeham law and late-life mortality plateaus. Using Canadian mortality data, we investigate the empirical behaviour of each approach and highlight the roles of heterogeneity, extrinsic risk, and stochastic randomness. Furthermore, we develop parallel definitions of biological age under both approaches and analyse how subjective survival beliefs emerge from misspecified parameters. Our comparison of these two approaches provides actuarial insights into the natural foundations of Gompertz mortality and the interpretation of ageing, frailty and death.
\end{abstract}

\noindent\textbf{Keywords:} Gompertz law; Reliability theory; Vitality modelling; Mortality plateau; Biological age
\bigskip

%%%%%%%%%%%%%%%%%%%%%%
%%% Section 1: Introduction
%%%%%%%%%%%%%%%%%%%%%%

\section{Introduction}

% Background
Understanding human mortality is an important task in demography, public health, and actuarial science. Mortality modelling supports numerous essential tasks such as life insurance pricing, annuity valuation, and pension risk management. Beyond traditional actuarial applications, it also plays an important role in assessing personal health, guiding public policy on social security systems, and informing individual life cycle decisions. At its core, mortality modelling seeks not only to quantify the risk of death over the human lifespan, but also to illuminate the mechanisms that drive physiological deterioration and shape mortality trajectories. This dual focus motivates a closer examination of the foundations of adult mortality patterns and encourages renewed attention to their biological and systemic origins.

% Gompertz
An important advancement in understanding adult mortality is the Gompertz law. First introduced by Benjamin Gompertz in 1825 \citep{gompertz1825mortality}, it states that mortality increases approximately exponentially with age. {This exponential pattern provides a good description of mortality across a broad range of adult ages, typically from around age 30 onward, although a gradual deceleration emerges at the oldest ages.} 
{The Gompertz law} has been confirmed across many populations and species and has become a standard tool in mortality modelling, life insurance valuation, and pension mathematics. Although many statistical extensions have been developed from the Gompertz framework, an important question remains{: why does mortality follow a Gompertz pattern across adult ages in the first place?} Identifying the mechanisms that give rise to Gompertz mortality is essential for uncovering universal drivers of ageing and for understanding how biological deterioration manifests over time.

% Mechanism
A natural step toward explaining the Gompertz law is to understand ageing through interpretable and explainable biological mechanisms. Two broad approaches have been widely studied in this context. The first views mortality as the consequence of accumulating frailty or damage within the organism and is referred to as the reliability approach. The second treats mortality as the depletion of a latent survival capacity, known as the vitality approach. Although they differ in biological interpretation, both approaches can generate Gompertz mortality patterns under suitable assumptions. This shared ability to explain the exponential rise in adult mortality suggests that Gompertz patterns may reflect fundamental properties of natural ageing rather than the outcome of a purely mathematical law.

% Reliability
Reliability-based explanations of ageing originate from engineering studies of system failure. Early work such as \cite{barlow1965reliability} developed probabilistic models to describe how complex engineered systems deteriorate through the failure of redundant components. A key step toward mortality modelling in this direction was taken by \cite{GAVRILOV20053}, who modeled an organism as a system composed of multiple blocks built from parallel redundant elements. In their formulation, ageing arises from the gradual exhaustion of redundancy, which leads to an exponential acceleration in mortality and can reproduce Gompertz and late life plateaus. More recent developments in this direction include \cite{flietner2024unifying}, who formalized a network view through a non homogeneous Markov process and demonstrated that Gompertzian hazards can emerge from damage propagation on complex networks, and \cite{nielsen2024gompertz}, who proposed a stochastic reliability model in which subsystems experience irreversible failures and death occurs when any fatal configuration is reached. These formulations capture accelerating mortality with minimal assumptions and provide parsimonious explanations for Gompertz patterns in interconnected biological systems.

% Vitality
Vitality-based approaches offer a complementary view of ageing by modelling death as the moment when an individual's survival capacity is depleted. {This idea dates back to \cite{gompertz1825mortality}, who attributed death to two co-existing causes, namely ``chance, without previous disposition to death or deterioration'' and ``a deterioration, or an increased inability to withstand destruction.''} The earliest {quantitative} formulation is from \cite{strehler1960general}, who proposed that vitality declines over time and death occurs when the remaining vitality {falls to} a certain level. Building on this idea, \cite{anderson1992vitality} introduced the first stochastic vitality model, representing vitality as a Brownian motion with drift and treating death as a first passage event at zero. This framework has since been extended in various ways in demographic and biology research, such as \cite{li2009vitality} and \cite{sharrow2016quantifying}. In actuarial literature, \cite{shimizu2021why} and \cite{shimizu2023survival} advanced vitality-based modelling by introducing a diffusion-driven survival energy process that yields a closed-form mortality function. More recently, \cite{zhu2025mortality} developed a general framework that describes vitality dynamics through multiple components and showed how Gompertz laws and other mortality patterns arise from different specifications of these components.

% Question
Despite extensive research, the reliability-based and vitality-based approaches for mortality modelling have not yet been systematically and comparatively examined from a unified actuarial perspective. A unified understanding of the Gompertz mortality pattern is valuable for numerous actuarial applications that rely on understanding structural drivers of mortality, including heterogeneity in populations, biological ages, and subjective survival beliefs. This paper aims to address this gap by bringing together the ideas of reliability and vitality modelling, providing a comparative analysis of their mechanisms, and generating empirical insights that can guide future actuarial research on mortality and ageing.

% Contribution 1
The first contribution of this paper is providing a unified actuarial view of reliability-based and vitality-based approaches for mortality modelling. Although the two approaches originate from different ageing mechanisms, we present both using consistent actuarial notation and interpretation, which allows their common structure and fundamental differences to be seen more clearly. We also show how each approach generates the Gompertz law under natural assumptions and highlight the biological reasons that drive these patterns. This unified view provides an avenue to enrich actuarial thinking on the dynamics of ageing and mortality through mechanistic perspectives.

% Contribution 2
The second contribution is an empirical comparison of the two modelling approaches through calibrated simulations. Using Canadian mortality data, we estimate models based on both approaches and examine how each reproduces Gompertz mortality patterns. We then analyse how model components, such as the distribution of initial vitality or frailty, influence the resulting mortality curve and the level of mortality plateaus. We also extend the models to reproduce the Makeham law and compare the accuracy of the fitted models. Our results provide empirical evidence on how traditional mortality laws can be reproduced and extended by biological modelling approaches.

% Contribution 3
Our last contribution is to examine how each modelling approach informs actuarial concepts related to biological age and subjective survival beliefs. For both approaches, we develop parallel definitions of biological age and analyse how these definitions reflect differences in health state and remaining lifetime. We then discuss how subjective beliefs can arise from misspecified model parameters and prove that the two approaches imply different relationships between average health and life expectancy. These results highlight how the two modelling approaches can offer insights into the construction of biological age measures and the interpretation of subjective beliefs.

% Structure
The rest of the paper is organised as follows. Section \ref{sec:Model} presents the vitality-based and reliability-based modelling approaches using consistent actuarial notation. Section \ref{sec:Analysis} calibrates both approaches to Canadian mortality data and compares their empirical performance under various specifications. Section \ref{sec:BioAge} introduces the biological age measures implied by both models and discusses subjective survival beliefs under each approach. Section \ref{sec:Conclusion} concludes with possible directions for future work.

%%%%%%%%%%%%%%%%%%%%%%
%%% Section 2: Model
%%%%%%%%%%%%%%%%%%%%%%

\section{Actuarial Formulation of Vitality and Reliability Models}\label{sec:Model}

The objective of this section is to establish a consistent actuarial foundation for understanding and applying the vitality-based and reliability-based approaches for mortality modelling. We first present the vitality framework, then provide a corresponding formulation for the reliability framework, and conclude with a comparative discussion that links the two perspectives.

\subsection{Vitality Model}\label{sec:Model_Vitality}

The idea of using vitality to model mortality traces back to the classical view that each individual possesses a latent survival capacity that evolves over the life course \citep{strehler1960general}. In actuarial science, \cite{zhu2025mortality} introduced a flexible vitality-based modelling framework, in which mortality arises as the first passage time of a stochastic vitality process to zero. This formulation provides a tractable structure that links vitality dynamics with various mortality models.

Let $x_0$ be the initial age of the individual at time $t=0$.\footnote{For notational brevity, we may omit $x_0$ when no confusion is caused.} The vitality process $V(t)$ describes the remaining survival capacity at time $t\ge 0$ and is governed by the stochastic dynamics
\begin{align}\label{eq:gompertz_vitality}
    V(t) = V_0 - \int_0^t \mu_{x_0}(s)\, \diff s - \sigma W(t) - \sum_{i=1}^{N(t)} Z_i,
\end{align}
where
\begin{itemize}
    \item $V_0$ is the initial vitality, which may be fixed or drawn from a distribution to capture heterogeneity in population health,
    \item $\mu_{x_0}(t)$ is a deterministic function describing the rate of vitality depletion, playing a role parallel to the force of mortality in classical models,
    \item $\sigma W(t)$ is the diffusion component, with Brownian motion $W(t)$ and diffusion parameter $\sigma$, representing individual specific random fluctuations in health,
    \item $\sum_{i=1}^{N(t)} Z_i$ is the jump component that captures sudden vitality losses due to accidents, illnesses, or catastrophic shocks.
\end{itemize}
We define the occurrence of death as the first time $V(t)$ crosses zero. It follows that, if the individual is alive at time $t$ (aged $x_0+t$) with vitality $v>0$, then their future lifetime is
\[
T_{x_0+t}(v)
=
\inf\{\tau\ge 0 : V(t+\tau)\le 0 \mid V(t)=v \}.
\]
This first passage time representation makes the vitality model directly interpretable and links survival probabilities to the distribution of the hitting time of a stochastic process.

A key advantage of the vitality approach is its ability to recover classical mortality laws. When the diffusion and jump components are omitted and the initial vitality is set to $V_0 \sim \mathrm{Exp}(1)$, then the $t$-year survival probability at age $x_0$ becomes
\[
\Pr(T_{x_0}(V_0) > t) = \exp\left( -\int_0^t \mu_{x_0}(s)\diff s \right).
\]
This expression coincides with the survival function implied by any force of mortality $\mu_{x_0}(t)$. Thus, any specific mortality law defined through $\mu_{x_0}(t)$ can be viewed as a special case of the vitality framework with the corresponding deterministic depletion rate. For example, setting $\mu_{x_0}(t)=b\, c^{t}$ yields the classical Gompertz law.\footnote{Other specifications of equation \eqref{eq:gompertz_vitality} can generate various mortality patterns. For instance, taking $V_0$ to follow a Gompertz distribution with a constant depletion rate also recovers the Gompertz law. We refer readers to \cite{zhu2025mortality} for additional examples and discussions, including death plateau and dynamic mortality models.}

We end this subsection with a short discussion on how the vitality-based approach provides several useful features for mortality modelling. Heterogeneity in health conditions across individuals in a population or a portfolio of insureds can be captured by the distributional choice of the initial vitality $V_0$. Both intrinsic and extrinsic sources of mortality can also be incorporated, through the diffusion term for individual health trajectories and the jump component for accidental deaths and large health shocks. These features allow the model to reproduce a wide range of mortality behaviours, including the Gompertz mortality curve and late-life mortality plateau, while maintaining an intuitive interpretation of survival dynamics.

\subsection{Reliability Model}\label{sec:Model_Reliability}

The reliability model rooted in the classical engineering idea that mechanical systems consist of redundant components that deteriorate and eventually fail. From an ageing perspective, such system can be viewed as an organism composed of multiple subsystems that accumulate failures over time \citep{gavrilov2005reliability}. More recently, in the context of mortality, \cite{nielsen2024gompertz} showed that the Gompertz law emerges naturally from subsystem failures in an interconnected biological system. In this subsection, we present an actuarial formulation of this framework and highlight several extensions.

\begin{figure}[ht!]
\begin{tikzpicture}[
    state/.style={draw, rounded corners=3pt, minimum size=4cm, inner sep=9pt},
    arrow/.style={-Stealth, very thick, shorten >=12pt, shorten <=12pt},
    cell/.style={minimum size=6mm, inner sep=0pt, outer sep=0pt,
                 draw=gray!40, line width=0.2pt},   % ← thin borders
    blackcell/.style={cell, fill=black, draw=none},
    redcell/.style={cell, fill=red!90!black, draw=none}
]

% --- Three states ---
\node[state] (A) at (0,0) {};
\node[state] (B) at (5,0) {};
\node[state] (C) at (11.9,0) {};
\node[] (D) at (8.5,0){\LARGE $\cdots$};

% Labels
\node[above=0.1cm of A]  {\large State $F_0$};
\node[above=0.1cm of B]  {\large State $F_0+1$};
\node[above=0.1cm of C]  {\large State $N$};
\node[above=1.9cm of D]  { };

% Arrows
\draw[arrow] (A) -> (B);
\draw[arrow] (B) -> (D);
\draw[arrow] (D) -> (C);

%\draw[arrow] (B) -> ++(5,0) node[midway,above=0pt,scale=2.5] {$\cdots$};
%\draw[arrow] (B) ++(5,0) -> (C);

% --- Macro to draw 6×6 grid ---
\newcommand{\drawsix}[4][]{% #1=extra style, #2=node, #3=black indices, #4=red index (0=none)
  \foreach \x in {0,...,5} \foreach \y in {0,...,5} {
      \coordinate (p) at ([shift={(-1.6cm + 0.64cm*\x, -1.6cm + 0.64cm*\y)}]#2.center);
      \node[cell, #1] at (p) {};
  }
  \foreach \i in {#3} {
      \pgfmathtruncatemacro{\x}{mod(\i,6)}
      \pgfmathtruncatemacro{\y}{\i/6}
      \coordinate (p) at ([shift={(-1.6cm + 0.64cm*\x, -1.6cm + 0.64cm*\y)}]#2.center);
      \node[blackcell] at (p) {};
  }
  \ifnum#4>0
      \pgfmathtruncatemacro{\x}{mod(#4,6)}
      \pgfmathtruncatemacro{\y}{#4/6}
      \coordinate (p) at ([shift={(-1.6cm + 0.64cm*\x, -1.6cm + 0.64cm*\y)}]#2.center);
      \node[redcell] at (p) {};
  \fi
}

% === Draw the three grids ===
% State A: 5 random black cells
\drawsix{A}{7,12,19,28,33}{0}

% State B: same 5 black + one red
\drawsix{B}{7,12,19,28,33}{17}

% State C: fully black (k=36)
\drawsix[fill=black]{C}{}{0}

\end{tikzpicture}
\caption{A multi-state structure for the subsystem failure process $F(t) \in \{F_0, F_0+1, \ldots, N\}$.}
\label{fig:gompertz_model}
\end{figure}
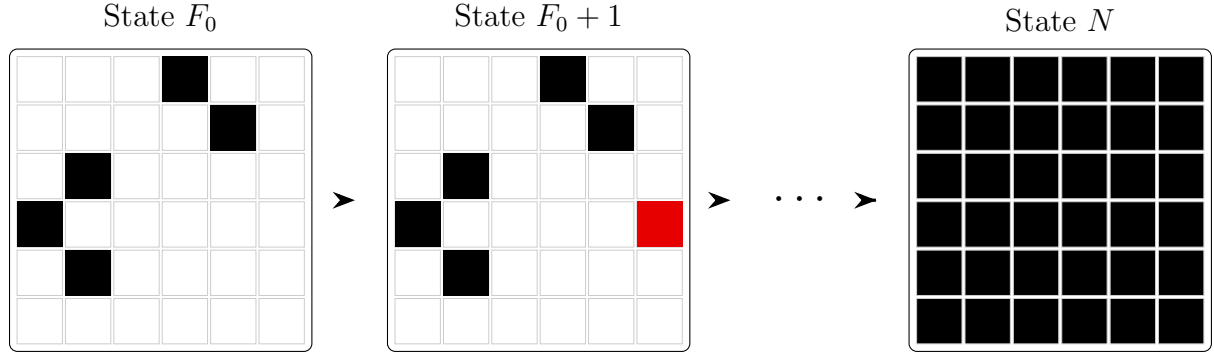

% Construction
Assume an organism consists of $N$ subsystems. At time $t$, let $F(t)$ denote the number of dysfunctional subsystems, with initial value $F(0)=F_0$ at time $t=0$ (age $x_0$). The process $F(t)$ evolves as a multi-state model on the state set $\{F_0, F_0+1, \ldots, N\}$. We then use standard actuarial notation
\[
{}_tp_{x_0+s}^{ij} = \Pr(F(s+t)=j \mid F(s)=i)
\]
for transition probabilities and define transition intensities as
\[
\lambda_x^{ij} = \lim_{h\to 0^+} \frac{{}_hp_x^{ij}}{h}, \qquad i\ne j.
\]
We assume the usual Markov and regularity conditions for multi-state models \citep{dickson2020actuarial}. Figure~\ref{fig:gompertz_model} provides a graphical illustration of this formulation.

% Transtion
The transition from $F(t)=k$ to $k+1$ (i.e., the failure rate) depends on two factors: (1) the number of remaining functioning subsystems $N-k$, and (2) the number of dysfunctional subsystems $k$. The first factor is natural since it represents how many subsystems are still available to fail, while the second factor arises from the assumption that dysfunctional subsystems may accelerate future failures due to inter-dependence among biological components \citep{nielsen2024gompertz}. To capture both effects, the transition intensity from state $k$ to $k+1$ is defined as
\begin{align*}
    \lambda_x^{k,k+1} = r\, k (N-k), \qquad k \ge F_0,
\end{align*}
where $r$ is a positive constant. Note that this transition intensity is age-independent conditional on $F(t)=k$ at time $t$. When $N$ is large, the law of large numbers yields a deterministic approximation for $F(t)$:
\begin{align}\label{eq:F_approx}
    F(t) \approx \frac{N F_0}{F_0 + (N-F_0)e^{-rNt}}.
\end{align}
This function is in a logistic form and thus suggests an accelerating accumulation of failures over time.

% Mortality
To resemble the Gompertz law, \cite{nielsen2024gompertz} assume that the force of mortality at time $t$ is proportional to the fraction of dysfunctional subsystems; that is, $\mu(t) \propto {F(t)}/{N}$. Substituting equation \eqref{eq:F_approx}, we obtain
\begin{align}\label{eq:gompertz_mu}
\mu(t) = \frac{\kappa F_0}{F_0 + (N-F_0)e^{-rNt}},
\end{align}
where $\kappa$ is a positive constant. It follows that the $t$-year survival probability at age $x_0$ is
\begin{align*}
    \Pr(T_{x_0}(F_0) > t) &= \left( 1 + \frac{F_0}{N} \left(e^{rNt} - 1\right)  \right)^{-\frac{\kappa}{rN}},
\end{align*}
where $T_{x_0}(F_0)$ is the future lifetime of an individual aged $x_0$ with initial value $F_0$. When $t$ is small and $N \gg F_0$, equation~\eqref{eq:gompertz_mu} is dominated by its exponential term and can be approximated as
\begin{align} \label{eq:gompertz_reliability_mu}
\mu(t) \approx \frac{\kappa F_0}{N} e^{rNt}.
\end{align}
Note that equation \eqref{eq:gompertz_reliability_mu} matches the Gompertz law $\mu(t) = b\, c^{t}$, with the re-parametrization $b = {\kappa F_0}/{N}$ and $c = e^{rN}$. In other words, equation \eqref{eq:gompertz_mu} produces Gompertzian mortality during most of adulthood, when $t$ is relatively small. As $t$ becomes large, the force of mortality $\mu(t)$ converges to the constant $\kappa$, producing a mortality plateau at extreme old ages.

We now offer several remarks that extend the reliability model presented in \cite{nielsen2024gompertz}. First, in gerontology, a widely used health metric is the \emph{frailty index}, defined as the proportion of observed health deficits for an individual \citep{mitnitski2001accumulation}. It is natural to interpret ${F(t)}/{N} \in [0,1]$ as a frailty index. Empirical studies have shown a strong linear relationship between mortality and the frailty index on the log scale \citep{mitnitski2002mortality}; that is, $\ln \mu(t) \propto \ln ({F(t)}/{N})$. Equation \eqref{eq:gompertz_mu} suggests using a logit transformation instead, such as considering $\logit \mu(t) \propto \logit ({F(t)}/{N})$, if $F(t)$ follows equation \eqref{eq:F_approx} and $N$ is large, then
\[
\logit \mu(t) = \kappa \ln\!\left(\frac{F_0}{\,N - F_0\,}\right) + (\kappa r N)\, t.
\]
This directly leads to a logit-transformed Gompertz law, where $\kappa \ln\!\left(\frac{F_0}{N-F_0}\right)$ is the intercept term and $\kappa r N$ is the slope term.

Second, similar to the vitality model, the reliability model accommodates heterogeneity naturally. One may treat the initial frailty level $F_0$ as a random variable, $F_0 \sim \mathbb{F}$, to reflect variation in baseline health across individuals in a population or an insured portfolio. In addition, instead of assuming either $\mu(t) \propto {F(t)}/{N}$ or $\logit \mu(t) \propto \logit {F(t)}/{N}$, one may consider a more general linear logit extension,
\[
\logit \mu(t) = \kappa\, \logit \frac{F(t)}{N} + \beta,
\]
where $\beta$ serves as a Makeham-type term for age-independent mortality. Lastly, we provide further technical details about the reliability model in Appendix \ref{app:gompertz_reliability}, including the ordinary differential equations and the approximation equations of $\mathbb{E}[F(t)|F_0]$ (used in equation \eqref{eq:gompertz_mu}) and $\text{Var}(F(t)|F_0)$, and a complete multi-state model representation of $F(t)$.

\subsection{Summary and Comparison}\label{sec:Model_Comparison}

% Similarity
Although the vitality model and the reliability model describe ageing through different biological mechanisms, they share several structural similarities. Both of them model mortality as the outcome of a latent ageing process, both accommodate heterogeneity and random shocks, and both can reproduce the Gompertz law. Their differences lie primarily in how the ageing process is formulated and how death is defined. Table \ref{tab:comparison} summarizes the key similarities and differences of these two approaches.

% Difference
From a modelling perspective, the reliability model treats ageing as the accumulation of subsystem failures. Its logistic growth structure naturally produces exponential mortality increase at mid-life and deceleration at extreme ages. In contrast, the vitality model views ageing as the depletion of survival capacity. Death is defined as the first passage of vitality to zero, which provides an intuitive representation of individual health trajectories.

\begin{table}[ht!]
\centering
\small
\begin{tabular}{  m{3cm}  m{5.7cm}  m{5.7cm} }
\toprule
\textbf{Feature} & 
\textbf{Reliability Model} & 
\textbf{Vitality Model} \\
\midrule

Death criterion & 
Force of mortality linked to the frailty level, e.g., $\mu(t) \propto F(t)/N$ & 
Death occurs when vitality reaches zero, i.e., $V(t)=0$ \\[10pt]

Ageing mechanism & 
Accumulation of frailty, e.g., transition rate $\lambda = r\,F(t)(N-F(t))$ & 
Depletion of vitality, e.g., mortality rate $\mu(t) = b\, c^t$ \\[10pt]

Heterogeneity & 
Initial frailty is given by $F(0)=F_0$ or specified as $F(0)\sim\mathbb{F}$ & 
Initial vitality is specified as $V_0 \sim \mathbb{F}$ \\[10pt]

Gompertz law &
Exponential behaviour of $\mu(t)$ in equation \eqref{eq:gompertz_mu} when $t$ is small &
Setting $V_0 \sim \mathrm{Exp}(1)$ and choosing $\mu(t)$ of Gompertz form \\[10pt]

Mortality plateau &
Convergence of $\mu(t)$ in equation \eqref{eq:gompertz_mu} when $t$ is large &
Setting $V_0 \sim \mathrm{Pareto}$ or $\mathrm{Gamma}$ \\[10pt]

Stochastic trajectory & 
Variation in random failure timing for $F(t)$ & 
Random fluctuation from the diffusion term $\sigma W(t)$ \\[8pt]

Extrinsic risk  & 
Additive Makeham-type term in the relationship between $\mu(t)$ and $F(t)$ & 
Sudden vitality loss and death from the jump component $\sum_{i=1}^{N(t)} Z_i$\\[7pt]

\bottomrule
\end{tabular}
\caption{Feature comparison between the vitality model and reliability model.}
\label{tab:comparison}
\end{table}

%%%%%%%%%%%%%%%%%%%%%
%%% Section 2: Analysis
%%%%%%%%%%%%%%%%%%%%%

\section{Empirical Analysis of Vitality and Reliability Models} \label{sec:Analysis}

This section examines how the vitality and reliability models behave when calibrated to empirical mortality data and evaluated through simulations. Using Canadian {female} mortality rates ages 30--{110} in year 2019 from the HMD,\footnote{HMD. Human Mortality Database. Max Planck Institute for Demographic Research (Germany), University of California, Berkeley (USA), and French Institute for Demographic Studies (France). Available at www.mortality.org (data downloaded on November 11, 2025).} we estimate the parameters of three models: the Gompertz law, the reliability model, and the vitality model. {The HMD reports single-year mortality rates up to age 110, and we calibrate all three models over this full range.} All parameters are calibrated by minimizing the relative squared error (RSE) between the model-implied values and the observed rates. The baseline specifications of the three models are provided as follows.

For the Gompertz law, we consider the standard form
\[
\mu_g(t)=b_g c_g^{\,t},
\]
where $b_g$ and $c_g$ are parameters to be estimated. For the reliability model, the number of total subsystems is set to $N=10^6$ \citep{nielsen2024gompertz}, which ensures that the approximation in equation \eqref{eq:gompertz_mu} holds. The parameters to be estimated from equation \eqref{eq:gompertz_mu} are $F_0$, $r$, and $\kappa$. To facilitate comparison with the Gompertz law, we report the transformed parameters $c_r=\exp(rN)$ and $b_r=\kappa{F_0}/{N}$, which are directly comparable to $b_g$ and $c_g$; that is, we write $\mu_r(t) = b_r c_r^{\,t}$.

For the vitality model, recall that if $V_0 \sim \mathrm{Exp}(1)$ with no diffusion and jump components, then the Gompertz law can be explicitly recovered by setting the depletion rate to $\mu_v(t)=b_vc_v^{\,t}$. We choose to modify this specification in order to align with the specification of the reliability model. In particular, we assume that the initial vitality $V_0$ follows a Pareto Type II distribution with shape parameter $\alpha$ and scale parameter $\alpha-1$. This choice is supported by (1) the number of parameters to be estimated is the same as the reliability model, (2) it produces mortality plateau similar to the reliability model, and (3) $\mathbb{E}[V_0]$ is the same as assuming $V_0 \sim \mathrm{Exp}(1)$ (i.e., $\mathbb{E}[V_0]=1$).\footnote{Since the scale parameter is chosen as the shape parameter $\alpha$ minus 1 in order to keep $\mathbb{E}[V_0]=1$, the shape of the resulting Pareto Type II distribution is not very sensitive to changes in $\alpha$. For example, increasing $\alpha$ from 10 to 100 reduces the variance $\frac{\alpha}{\alpha - 2}$ from 1.25 to 1.0204.}

\subsection{Baseline results} \label{sec:Analysis_Baseline}

Table~\ref{tab:gompertz_parameters} reports the estimated parameters and RSE values for the fitted Gompertz, reliability and vitality models. The reliability and vitality models produce remarkably similar estimates for the growth parameter $c$ and the level parameter $b$, despite their different ageing mechanisms and the additional parameters $F_0=137.05$ and $\alpha=10.97$. The estimated $F_0$ is small relative to $N=10^6$, which is consistent with the interpretation that only a small fraction of subsystems are dysfunctional at the starting age. For the vitality model, the large value of $\alpha$ indicates a light-tailed distribution for initial vitality. Relative to these two models, the Gompertz law has a noticeably higher estimated $b$ and a slightly lower estimated $c$, implying a higher overall mortality level and a slower growth rate. The differences in goodness-of-fit are reflected in the models' RSE, where the reliability and vitality models substantially outperform the Gompertz law.

\begin{table}[ht!]
    \centering
    \begin{tabular}{c|ccc}
    \toprule    
    Parameter   & Gompertz Law & Gompertz-Reliability Model & Gompertz-Vitality Model \\
    \midrule
        $b$        & $3.2298\times 10^{-4}$ & $1.6951\times 10^{-4}$ & $1.5430\times 10^{-4}$ \\
        $c$        & $1.1022$               & $1.1194$               & $1.1194$               \\
        $F_0$      & --                     & $137.0458$             & --                     \\
        $\alpha$   & --                     & --                     & $10.9706$             \\
    \midrule
        RSE        & $1.2827$               & $0.7559$               & $0.7559$           \\
    \bottomrule
    \end{tabular}
    \caption{Estimated parameters and relative squared error (RSE) from the fitted Gompertz law and Gompertz-specification of the reliability and vitality models.}
    \label{tab:gompertz_parameters}
\end{table}

\begin{figure}[ht!]
    \centering
    \includegraphics[width=0.75\linewidth]{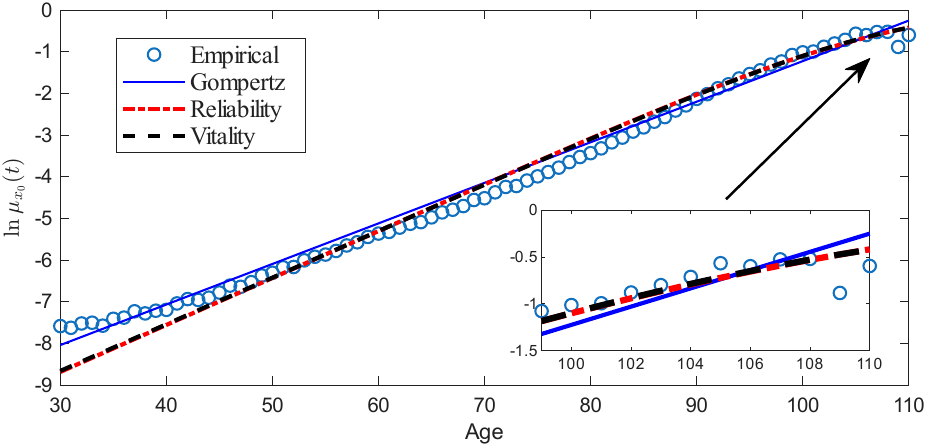}
    \caption{Observed mortality rates compared to the fitted Gompertz, reliability, and vitality models under the baseline specifications.}
    \label{fig:gompertz_benchmark}
\end{figure}

Figure~\ref{fig:gompertz_benchmark} displays the fitted mortality curves and highlights several visual differences across the models. In the main panel, all three curves track the observed mortality pattern well over ages 45-105. At younger ages 30-45, the Gompertz curve lies closer to the observed values, while both the reliability and vitality models noticeably underestimate mortality. {Over middle ages 65-90, the observed mortality values form a curve that is slightly below a straight line, and none of the three models fully captures this mid-age concavity.} In the inset panel, for ages above 99, the Gompertz model shows no sign of mortality plateau, whereas the other two models exhibit a modest extreme-age flattening. However, the plateau effect appears to be weaker than the empirical pattern shown. These observations indicate that none of the baseline specifications fully captures the empirical features observed. In the subsequent subsections, we examine model refinements based on the baseline specifications to address these issues.

Figure~\ref{fig:gompertz_benchmark} displays the fitted mortality curves and highlights several visual differences across the models. In the main panel, all three curves track the observed mortality pattern well over ages 45-105. At younger ages 30-45, the Gompertz curve lies closer to the observed values, while both the reliability and vitality models noticeably underestimate mortality. {Over the intermediate ages of roughly 65 to 90, the observed log mortality curve bends slightly below all three fitted curves, and none of the models captures this mid-age concavity.} In the inset panel, for ages above 99, the Gompertz model shows no sign of mortality plateau, whereas the other two models exhibit a modest extreme-age flattening. However, the plateau effect appears to be weaker than the empirical pattern shown. These observations indicate that none of the baseline specifications fully captures the empirical features observed. {Re-estimating the three models using other calendar years' data shows the same concavity over ages 65 to 90, further indicating that additional model flexibility is needed.} In the subsequent subsections, we examine model refinements based on the baseline specifications to address these issues.

{Appendix \ref{app_male} displays the calibration results using male mortality data. As expected, both the reliability and vitality models continue to outperform the Gompertz law. However, the improvement in goodness-of-fit is less substantial than in the female case, which we attribute to the less pronounced mortality plateau in the male data over the observed age range. This also explains why the estimated Gompertz parameters $b$ and $c$ are more similar across the three models. For the vitality model, the large estimate of the shape parameter $\alpha$ suggests that the initial vitality distribution $V_0$ is close to the exponential distribution.}

\subsection{Heterogeneity and mortality plateau} \label{sec:Analysis_Heterogeneity}

We first examine how heterogeneity in the initial vitality or frailty affects the resulting mortality pattern. Specifically, we assume that $V_0$ under the vitality model and $F_0$ under the reliability model follow either a Pareto Type II distribution (used in the vitality model's baseline specification) or a Gamma distribution. Our focus is on how the shape parameter $\alpha$ influences mortality curve. The associated scale parameter is set to match the baseline value $F_0=137.05$ for reliability or $\mathbb{E}[V_0]=1$ for vitality. We remark that a Pareto Type II distribution for $V_0$ can recover the Gamma-Gompertz mortality law, and Gamma distributions are widely used in describing frailty index behaviour in gerontology \citep{mitnitski2001accumulation, rockwood2004changes}.

\begin{figure}[ht!]
    \centering
    \includegraphics[width=0.80\linewidth]{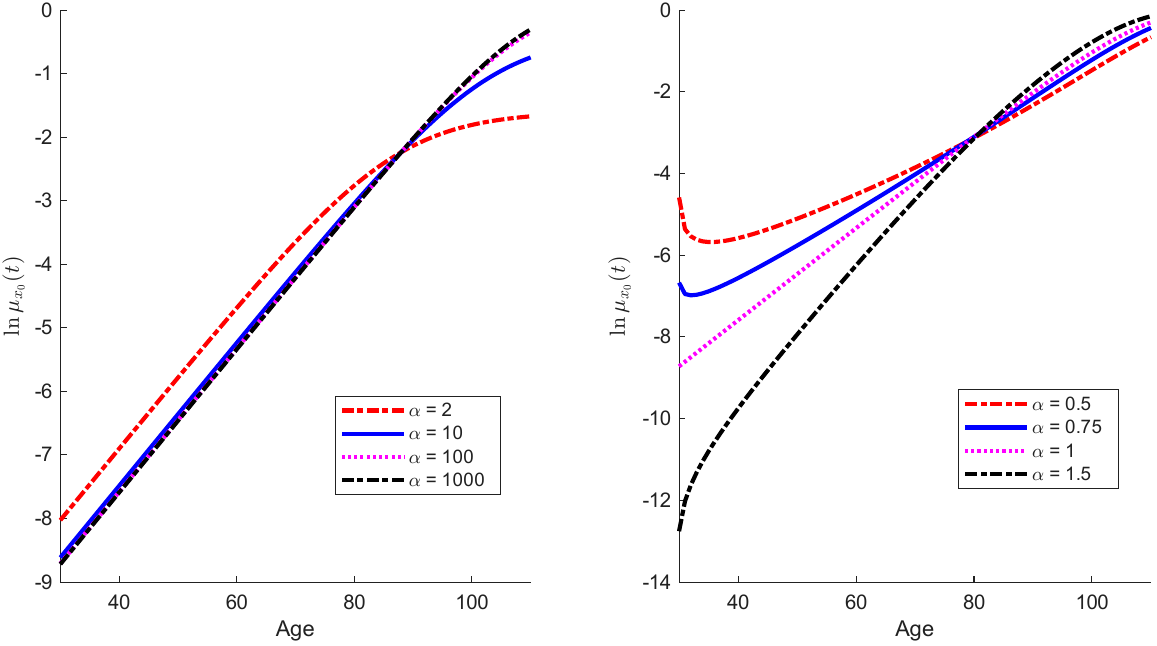}
    \caption{Vitality model under different distributions of $V_0$ with varying shape parameter $\alpha$: Pareto Type II distribution (left) and Gamma distribution (right).}
    \label{fig:gompertz_vitality_hetero}
\end{figure}

Figure~\ref{fig:gompertz_vitality_hetero} shows that the vitality model is responsive to the distribution of $V_0$ and parameter $\alpha$. Under the Pareto specification, smaller values of $\alpha$ imply heavier tails and a larger proportion of individuals with high initial vitality. This will slow down the increase in mortality at advanced ages and results in a stronger effect of mortality plateau.\footnote{
When $V_0 \sim \text{Pareto Type II}$ with shape parameter $\alpha$ and scale parameter $\alpha - 1$, and the depletion rate is specified as $b c^t$, we have
\[
    \mu(t) = \frac{\alpha b c^t}{ (\alpha - 1) + \frac{b}{\ln c}\left(c^t - 1\right)}
    \quad \text{and} \quad
    \mu'(t) = \mu(t) \left( \ln c - \frac{\mu(t)}{\alpha} \right).
\]
As $\mu(t) > 0$, $\mu'(t)$ is an increasing function of $\alpha$, implying that smaller values of $\alpha$ will lead to a stronger mortality plateau.} 
Under the Gamma specification, the resulting mortality curve around both advanced ages and younger ages can be significantly reshaped by $\alpha$. We thus conclude that the mortality curve implied by the vitality model responds strongly to the behaviour of the distribution of initial vitality.

\begin{figure}[ht!]
    \centering
    \includegraphics[width=0.85\linewidth]{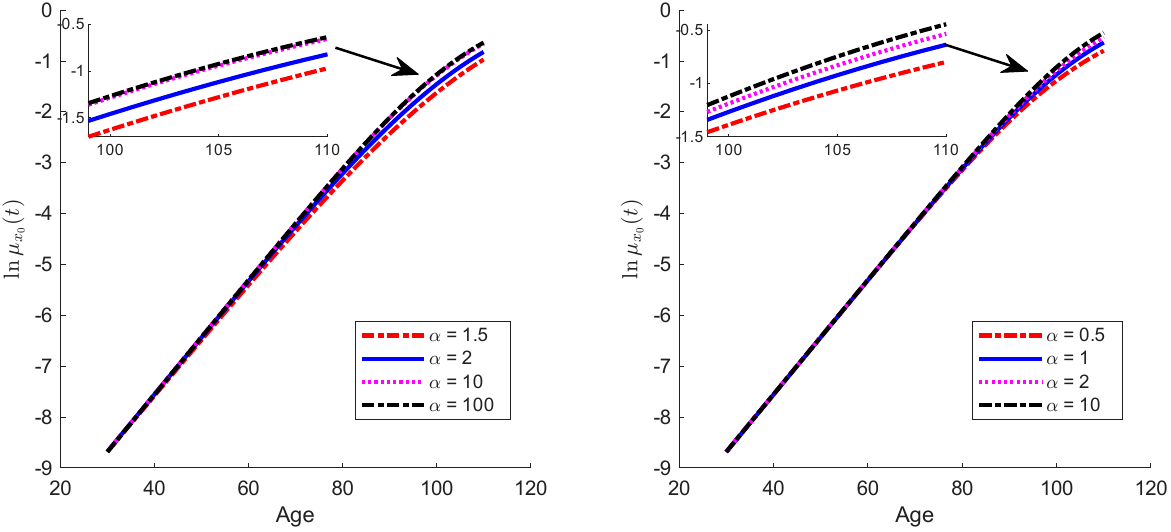}
    \caption{Reliability model under different distributions of $V_0$ with varying shape parameter $\alpha$: Pareto Type II distribution (left) and Gamma distribution (right).}
    \label{fig:gompertz_reliability_hetero}
\end{figure}

Figure~\ref{fig:gompertz_reliability_hetero} displays the same heterogeneity analysis for the reliability model. In contrast to the vitality case, the resulting mortality curves are largely insensitive to the distribution of $F_0$ or the choice of $\alpha$. The curves remain tightly clustered over most ages, except advanced ages as shown in the inset panels. This insensitivity can be loosely explained by the ageing mechanism of the reliability model. Since $N \gg F_0$ and $F(t)$ evolves according to a logistic form, changes in the distribution of $F_0$ have relatively limited impact on the force of mortality once the exponential growth dominates.

\subsection{Makeham law} \label{sec:Analysis_Makeham}

The Makeham law extends the Gompertz law by introducing an age-independent term for the force of mortality, which we specify as 
\[
\mu_m(t) = \beta_m + b_m c_m^t,
\]
where $\beta_m$ is the age-independent Makeham term. {This term can also be incorporated into the reliability and vitality models, but is accommodated differently in each. For the vitality model, it enters naturally through the depletion rate as $\mu_v(t)=\beta_v + b_vc_v^{\,t}$. For the reliability model, it does not arise naturally from the subsystem-failure mechanism but is instead introduced as an extrinsic term independent of subsystem failures, resulting in $\mu_r(t) = \beta_r + b_r c_r^{\,t}$ with $b_r=\kappa{F_0}/{N}$, and $c_r=\exp(rN)$.} We re-estimate all model parameters under this Makeham extension, while keeping the same baseline assumptions for $V_0$ and $F_0$.

\begin{table}[ht!]
    \centering
    \begin{tabular}{c|ccc}
    \toprule
    Parameter   & Makeham Law & Makeham-Reliability Model & Makeham-Vitality Model \\
         \midrule
      $\beta$   &  $1.6220\times 10^{-5}$ & $7.0423\times 10^{-4}$  & $5.8984\times 10^{-4}$ \\ 
      $b$       &  $3.1940\times 10^{-4}$ & $4.7081\times 10^{-5}$ & $4.0975\times 10^{-5}$ \\
      $c$       &  $1.1024$ & $1.1442$ & $1.1438$ \\
      $F_0$     &  -- & $54.1074$ & -- \\
      $\alpha$  &  -- & -- & $6.5226$ \\
      \midrule
      RSE       &  $1.2826$ & $0.4440$ & $0.4441$ \\
      \bottomrule
    \end{tabular}
    \caption{Estimated parameters and relative squared error (RSE) for the fitted Makeham Law and Makeham-specification of the reliability and vitality models.}
    \label{tab:gompertz_makeham}
\end{table}

\begin{figure}[ht!]
    \centering
    \includegraphics[width=0.75\linewidth]{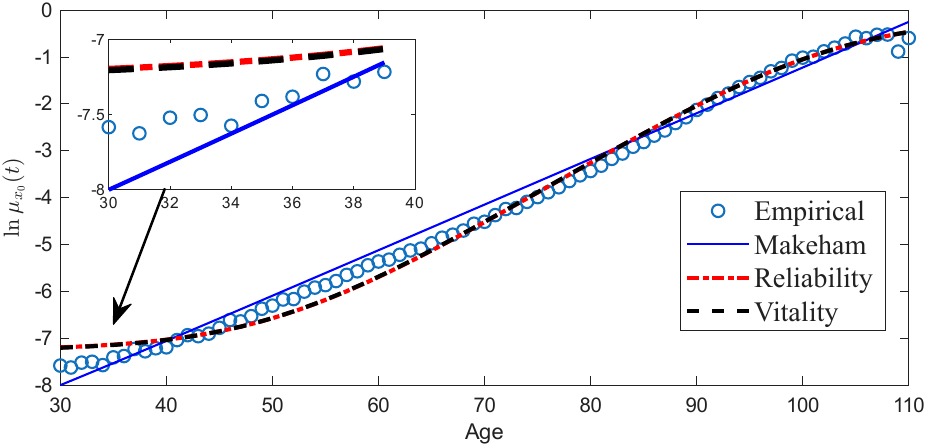}
    \caption{Fitted mortality curves under the Makeham Law and the extension from the reliability and vitality models.}
    \label{fig:gompertz_makeham}
\end{figure}

Table~\ref{tab:gompertz_makeham} summarizes the parameter estimates and fits, and Figure~\ref{fig:gompertz_makeham} displays the fitted mortality curves. The Makeham law has a negligible estimate of $\beta$. In contrast, both the reliability and vitality models estimate a non-zero $\beta$ of similar magnitude, which significantly improves the model fit. As shown in Figure~\ref{fig:gompertz_makeham}, the Makeham specification of the two models capture mortality at ages 30-45 more accurately while preserving the mid- and late-life performance as in the baseline cases with Gompertz specification. The RSE values decrease sharply to 0.3370 for the reliability model and 0.3264 for the vitality model, indicating that the additional Makeham term materially improves the goodness-of-fit of both models relative to the baseline specifications.

Lastly, for the vitality model, we note that the Makeham effect can be generated not only from modifying the depletion rate but also through other specifications of equation \eqref{eq:gompertz_vitality}. For instance, a fatal jump process ($\sum_{i=1}^{N(t)} Z_i$) with an appropriate diffusion term ($\sigma W(t)$) can reproduce the Makeham effect. When $V_0 \sim \mathrm{Exp}(1)$, introducing fatal jumps with constant intensity $\beta$ yields a survival function identical to that implied by adding a Makeham term to the depletion rate $\mu_v(t)$.\footnote{Let $T_{x_0}(V_0)$ be the future lifetime with initial vitality $V_0 \sim \mathrm{Exp}(1)$, and $\lambda_{x_0}(t)$ be the intensity of fatal jumps (i.e., $Z_i=\infty, \forall i$) at time $t$. With $\mu_{x_0}(t)$ being the depletion rate and $\sigma = 0$ from equation \eqref{eq:gompertz_vitality}, we have
\[
\Pr(T_{x_0}(V_0)>t)
= \exp\!\left(-\int_0^t \lambda_{x_0}(s)\,\mathrm{d}s\right)
  \times
  \exp\!\left(-\int_0^t \mu_{x_0}(s)\,\mathrm{d}s\right).
\]
If $\mu_{x_0}(t)$ follows the Gompertz law and $\lambda_{x_0}(t)=\beta$, then the resulting survival function coincides with the Makeham law.}

{The calibration results using male mortality data are presented in Appendix \ref{app_male}. As in the Gompertz case, the mortality plateau is less pronounced, which leads to closer parameter estimates across the three models and a less substantial improvement in goodness-of-fit.}

\subsection{Diffusion} \label{sec:Analysis_Diffusion}

We next examine the effect of including a diffusion term in the vitality model. Following \cite{zhu2025mortality}, we consider a benchmark volatility of $\hat{\sigma} = 9.8267\times 10^{-3}$ and evaluate mortality outcomes under different multiples of this value.

\begin{figure}[ht!]
    \centering
    \includegraphics[width=0.75\linewidth]{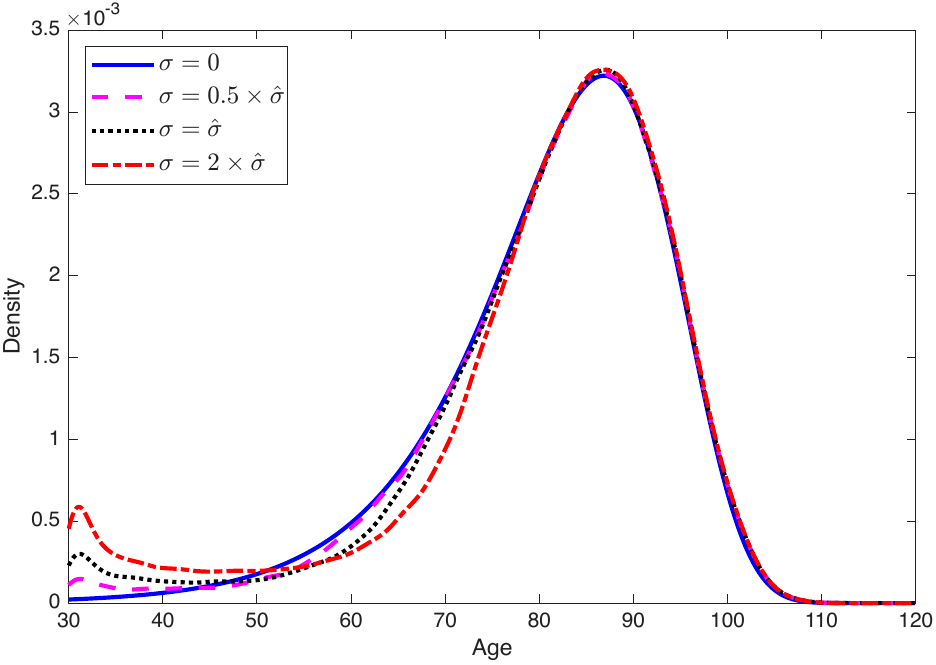}
    \caption{Effect of diffusion levels on the distribution of {the age at} death under the vitality model.}
    \label{fig:gompertz_diffusion}
\end{figure}

Figure~\ref{fig:gompertz_diffusion} displays the resulting {distribution of the age at death} for different choices of $\sigma$. As diffusion increases, the probability of experiencing an early death rises when initial vitality is low, leading to a small hump in the death density at younger ages (approximately ages 30-35). The overall shape of the distribution remains broadly similar across different choices of $\sigma$. These results suggest that diffusion mainly influences early-adult mortality while leaving mid- and late-life behaviour largely unchanged from a population-level {age-at-death distribution} perspective.

%%%%%%%%%%%%%%%%%%%%%
%%% Section 4: BioAge
%%%%%%%%%%%%%%%%%%%%%

\section{Biological Age and Ageing} \label{sec:BioAge}

This section examines how the ageing and mortality mechanisms under the reliability-based and vitality-based approaches relate to biological ageing and subjective mortality beliefs. One of the main advantages of these approaches is their ability to construct biological age. In contrast with chronological age, biological age aims to capture the underlying ageing process and reflect an individual’s biological condition. The construction of biological age has been widely studied in the biomedical literature \citep[][among many others]{nakamura1988assessment, hochschild1989improving, klemera2006new, kang2018biological, sluiskes2024accelerage}. We refer interested readers to the recent comprehensive review by \cite{bafei2023biomarkers}. Under our notation, the chronological age of an individual who has survived to time $t$ is $\text{CA} = x_0 + t$ regardless of the model considered.

{Two broad strategies can be used to translate an individual's health status into an equivalent biological age. The first is to match the individual to the population, by identifying the chronological age at which an average person is in the same condition, measured either by current health or by remaining lifetime, and reporting that age as the biological age. The second is to adjust the individual's own chronological age by the difference between their remaining lifetime and that of an average person of the same age. We develop concrete definitions based on these two strategies for each model in the following subsections.}

Before introducing the formal definitions of biological age, we specify the model settings used for illustration purposes and present representative individual trajectories of both vitality and frailty. Our analysis largely follows the baseline specifications from Section \ref{sec:Analysis}, but includes two additional modifications. For the reliability model, we assume that initial frailty $F_0$ follows a Gamma distribution with a shape parameter of 10 and a scale parameter of 0.1. For the vitality model, we include a diffusion term with volatility $\hat{\sigma} = 9.8267\times 10^{-3}$ \citep{zhu2025mortality}. 

\begin{figure}[ht!]
    \centering
    \includegraphics[width=0.85\linewidth]{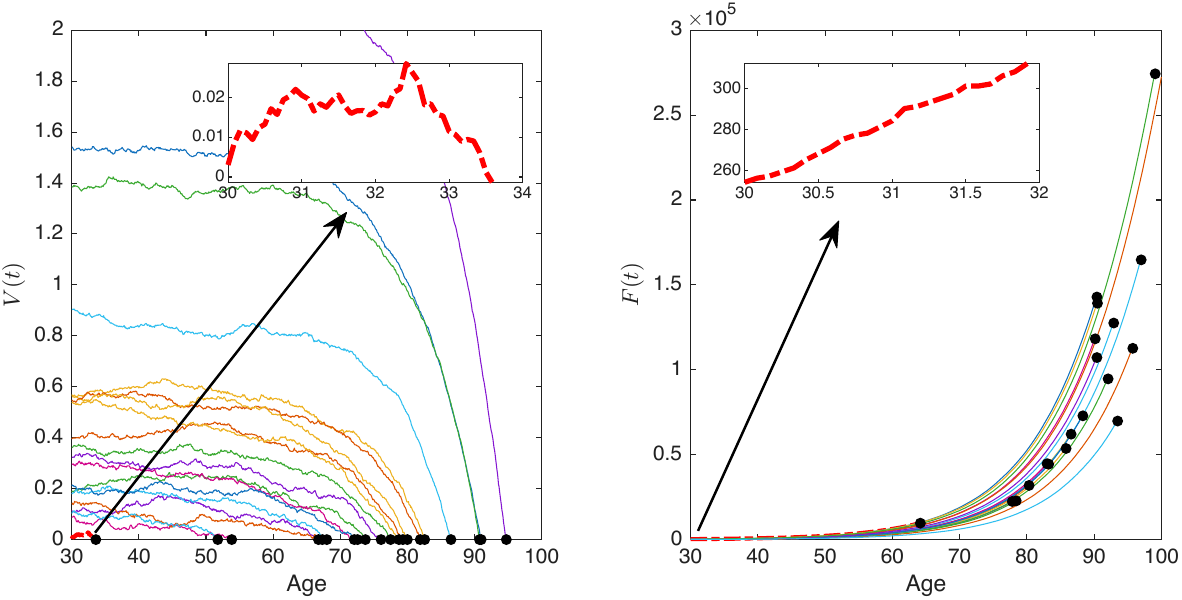}
    \caption{Simulated trajectories under the vitality model (left) and the reliability model (right). {Each curve is one individual's vitality $V(t)$ (left) or number of failed subsystems $F(t)$ (right), the solid dots mark the simulated ages at death, and the inset magnifies a single trajectory near the initial age.}}
    \label{fig:gompertz_trajectory}
\end{figure}

Figure~\ref{fig:gompertz_trajectory} shows simulated trajectories under the vitality and the reliability models. In the vitality model, each curve traces an individual's vitality $V(t)$ over time/age, with death occurring when $V(t)$ first reaches zero. The trajectories differ in their starting points according to $V_0$, share the same Gompertz depletion trend, and exhibit stochastic fluctuations due to the diffusion term. In the reliability model, each curve represents an individual's accumulation of subsystem failures $F(t)$ over time/age, with solid dots indicating simulated death times. The trajectories differ in their initial values of $F_0$ and reflect both the overall logistic trend and the randomness in the timing of failure events (see Appendix \ref{app:gompertz_reliability} for details). {We emphasise that the subscript in $V_0$ and $F_0$ refers to the starting time $t=0$, not the chronological age $0$, and the starting age is $x_0=30$.}

\subsection{Biological age under the reliability model} \label{sec:BioAge_Reliability}

The construction of biological age involves comparing an individual's current biological state with that of a reference individual in the population. Under the reliability model, a natural definition is to match the individual's current frailty level with the frailty trajectory of an ``average'' person. This idea has been used in \cite{nielsen2024gompertz} and is conceptually similar to the longevity‐risk‐adjusted age in \cite{milevsky2020calibrating}, where backward engineering is used to determine the implied age relative to a global average under the Makeham law. Based on this idea, we have following definition:
\begin{definition} \normalsize \label{def:BA_Back_Reliability}
Under the reliability model, for an individual $i$ with frailty level $F_i(t)$ at time $t$, the {backward health-matching} biological age is
\begin{align*}
    \text{BA}_{hm}(F_i(t))
    = x_0 + \inf \left\{ \tau : F_i(t)
      = \mathbb{E}\!\left[F(\tau;0,\mathbb{E}[F_0])\right] \right\},
\end{align*}
where $F(\tau; t, f)$ is an increasing function of $\tau$ for the frailty level at time $\tau$ for an individual whose frailty at time $t<\tau$ is $f$.
\end{definition}

If the approximation from equation \eqref{eq:F_approx} is used, then this biological age admits the closed-form expression:
\begin{align}\label{eq:gompertz_BA_reliability}
    \text{BA}_{hm}(F_i(t))
    = x_0 - \frac{1}{rN}
      \ln\!\left(
      \frac{\mathbb{E}[F_0]/F_i(t) - \mathbb{E}[F_0]/N}
           {1 - \mathbb{E}[F_0]/N}
      \right).
\end{align}
When $\mathbb{E}[F_0]\ll N$ and $t$ is small, this expression simplifies to
\begin{align*}
    \text{BA}_{hm}(F_i(t))
    \approx x_0 - \frac{1}{rN}\ln\!\left(\frac{\mathbb{E}[F_0]}{F_i(t)}\right)
    = \left(x_0 - \frac{1}{rN}\ln\!\frac{\mathbb{E}[F_0]}{N}\right)
      + \frac{1}{rN}\ln\!\left(\frac{F_i(t)}{N}\right),
\end{align*}
which is consistent with how the biological age and frailty index are linked in \cite{mitnitski2002frailty}. We note that the biological age from Definition \ref{def:BA_Back_Reliability} is independent of time $t$ or equivalently chronological age $x_0+t$.

We refer to Definition~\ref{def:BA_Back_Reliability} as the \emph{backward} method, since it determines biological age by matching the individual's frailty level to a reference point on the population frailty curve. {The name reflects that the method is based on the individual's current health status due to their past ageing, and we accordingly also refer to it as the \emph{health-matching} method, denoted $\text{BA}_{hm}$.} Alternatively, we also propose a \emph{forward} method that adjusts the individual's current chronological age by the difference in expected remaining lifetime between the individual and an average person at the same chronological age. {The name reflects that this method is instead based on the individual's projected future lifetime, and we accordingly also refer to it as the \emph{age-shifting} method, denoted $\text{BA}_{as}$.} The formal definition is given as follows:
\begin{definition} \normalsize \label{def:BA_Forw_Reliability}
Under the reliability model, for an individual $i$ with frailty level $F_i(t)$ at time $t$, the forward {age-shifting} biological age is
\begin{align}\label{eq:gompertz_BA_reliability2}
    \text{BA}_{as}(t, F_i(t))
    = x_0 + t
      + \mathbb{E}\!\left[
          T_{x_0+t}\!\left(\mathbb{E}[F(t;0,\mathbb{E}[F_0])]\right)
        \right]
      - \mathbb{E}\!\left[T_{x_0+t}(F_i(t))\right],
\end{align}
where $T_{x}(f)$ is the remaining lifetime of an individual aged $x$ with frailty level $f$, and $F(\tau; t, f)$ is an increasing function of $\tau$ for the frailty level at time $\tau$ for an individual whose frailty at time $t<\tau$ is $f$.
\end{definition}
\noindent
We note that, in equation \eqref{eq:gompertz_BA_reliability2}, the first expectation corresponds to the expected remaining lifetime of an average person, while the second corresponds to that of the individual. The difference between these two provides the adjustment to the individual's chronological age $x_0+t$ to reach the forward biological age.

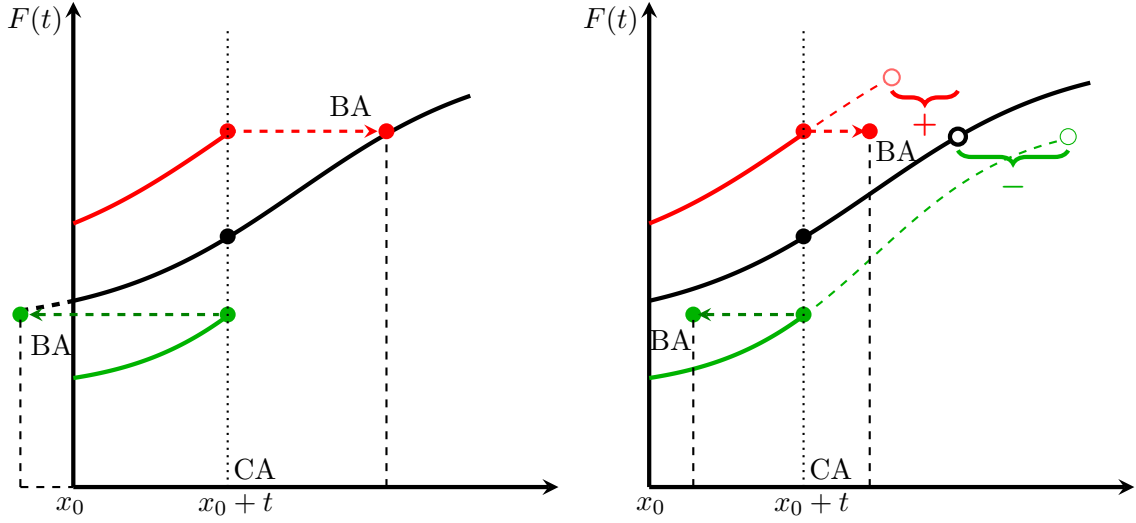
\begin{figure}[ht!]
\centering
\begin{tikzpicture}
\begin{groupplot}[
group style={
        group size=2 by 1,              % 2 columns, 1 row
        horizontal sep=1.2cm,           % gap between the two plots
        group name=plots,               % for referencing if needed
    },
    axis lines=left,
    ylabel style={at={(0.05,0.95)},rotate=90,anchor=south},
    xtick=\empty, ytick=\empty,
    xmin=0, xmax=110,
    ymin=0, ymax=90,
    width=8cm, height=8cm,
    clip=false,
    axis line style={ultra thick},
    title style={at={(0.5,1.05)},anchor=north}
]

%% ==================  Left Panel =================================
%% =================================================================
\nextgroupplot

\node[above left] at (axis cs:0,82) {$F(t)$};
\node[below left] at (axis cs:5,0) {$x_0$};

\addplot[ultra thick, black, domain=0:90, samples=200] 
    {50 / (1 + 10*exp(-0.045*x))+30};
\addplot[ultra thick, black, dashed, domain=-12:0, samples=200] 
    {50 / (1 + 10*exp(-0.045*x))+30};
\fill[black] (axis cs:35,46.5) circle (3pt);

%% Healthy Person
\addplot[ultra thick, green!70!black, domain=0:35, samples=200] 
     {50 / (1 + (450/20-1)*exp(-0.06*x))+18};

\draw[dashed, green!50!black, very thick, stealth-] (axis cs:-10,32) -- (axis cs:33, 32)
                     node[midway, above, sloped, black] {};

\fill[green!70!black] (axis cs:-12,32) circle (3pt);
\fill[green!70!black] (axis cs:35,32) circle (3pt);

\node[below right] at (axis cs:-12,30) {BA};
\node[below right] at (axis cs: 34,7) {CA};
\node[below right] at (axis cs: 26,0.8) {$x_0+t$};
\draw[dashed, thick] (axis cs:-12,0) -- (axis cs:-12, 32)
                     node[midway, above, sloped, black] {};
                     
\draw[dashed, thick] (axis cs:-12,0) -- (axis cs:0, 0)
                     node[midway, above, sloped, black] {};
                     
%% Less Healthy Person
\addplot[ultra thick, red, domain=0:35, samples=200] 
     {50 / (1 + (450/80-1)*exp(-0.045*x))+40};
\draw[dashed, red, very thick, -stealth] (axis cs:35,66) -- (axis cs:69, 66) node[midway, above, sloped, black] {};

\fill[red] (axis cs:35,66) circle (3pt);
\fill[red] (axis cs:71,66) circle (3pt);

\draw[dashed, thick] (axis cs:71,0) -- (axis cs:71, 65)
                     node[midway, above, sloped, black] {};

%\node[above left] at (axis cs:35,45) {CA};
\node[above left] at (axis cs:70,67) {BA};

\draw[dotted, thick] (axis cs:35,0) -- (axis cs:35, 85)
                     node[midway, above, sloped, black] {};

%% ================ Right Penal ======================================

\nextgroupplot
\node[above left] at (axis cs:0,82) {$F(t)$};
\node[below left] at (axis cs:6,0) {$x_0$};

\addplot[ultra thick, black, domain=0:100, samples=200] {50 / (1 + 10*exp(-0.045*x))+30};
\filldraw[black, fill=black!1, ultra thick] (axis cs:70, 65)circle(3pt);
\fill[black] (axis cs:35,46.5) circle (3pt);

%% Healthy Person
\addplot[ultra thick, green!70!black, domain=0:35, samples=200] 
     {50 / (1 + (450/20-1)*exp(-0.06*x))+18};
\addplot[thick, dashed, green!70!black, domain=35:95, samples=200] 
     {50 / (1 + (450/20-1)*exp(-0.06*x))+18};

\filldraw[green!70!black, fill=green!1] (axis cs:95,65) circle (3pt);
\fill[green!70!black] (axis cs:35,32) circle (3pt);

\draw[ultra thick,decorate,green!70!black, decoration={brace,amplitude=5pt,raise=-5pt,mirror}]
    (axis cs:70,60) -- node[below=0pt] {$\boldsymbol{-}$ } (axis cs:95,60);

% BA for healthy person
\fill[green!70!black] (axis cs:10,32) circle (3pt);
\node[below left] at (axis cs:12,31) {BA};

%\draw[ultra thick,decorate,green!70!black, decoration={brace,amplitude=5pt,raise=2pt,mirror}]    (axis cs:11,30) -- node[above=8pt] { } (axis cs:34,30);
\draw[dashed, green!50!black, very thick, stealth-] (axis cs:11,32) -- (axis cs:34,32)
                     node[midway, above, sloped, black] {};

\draw[dashed, thick] (axis cs:10,0) -- (axis cs:10, 31)
                     node[midway, above, sloped, black] {};

%% Less Healthy Person
\addplot[ultra thick, red, domain=0:35, samples=200] 
     {50 / (1 + (450/80-1)*exp(-0.045*x))+40};
\addplot[thick, dashed, red, domain=20:55, samples=200] 
     {50 / (1 + (450/80-1)*exp(-0.045*x))+40};
          
\fill[red] (axis cs:35,66) circle (3pt);
\filldraw[color = red!60,fill=red!1, thick] (axis cs:55,76) circle(3pt);
\fill[red] (axis cs:50,66) circle(3pt);
\node[below right] at (axis cs:49,66) {BA};

\draw[ultra thick,decorate,red, decoration={brace,amplitude=5pt,raise=1pt,mirror}]
    (axis cs:55,74) -- node[below=5pt] {$\boldsymbol{+}$ } (axis cs:70,74);
%\draw[ultra thick,decorate,red, decoration={brace,amplitude=5pt,raise=2pt,mirror}]
%    (axis cs:30,61) -- node[below=5pt] {} (axis cs:45,61);
\draw[dashed, red, very thick, -stealth] (axis cs:36,66) -- (axis cs:49,66)
                     node[midway, above, sloped, black] {};

\draw[dashed, thick] (axis cs:50,0) -- (axis cs:50, 65)
                     node[midway, above, sloped, black] {};

\draw[dotted, thick] (axis cs:35,0) -- (axis cs:35, 85)
                     node[midway, above, sloped, black] {};
                     
\node[below right] at (axis cs: 34,7) {CA};
\node[below right] at (axis cs: 26,0.8) {$x_0+t$};

\end{groupplot}
\end{tikzpicture}
\caption{Visual illustration of how biological age is determined under the reliability model using the backward {health-matching} method (left) and the forward {age-shifting} method (right). {The solid black curve is the average frailty trajectory and the green and red curves are a healthier and a frailer individual, respectively, with filled circles marking chronological age (CA) and biological age (BA). In the right panel, the open circles denote the projected end points of remaining lifetime, the dashed segments show the projected future frailty, and the $+$ and $-$ signs indicate whether the adjustment is added to or subtracted from CA to obtain BA.}}
\label{fig:BA_illustration_reliability}
\end{figure}

Figure \ref{fig:BA_illustration_reliability} provides a graphical illustration of Definitions \ref{def:BA_Back_Reliability} and \ref{def:BA_Forw_Reliability}. The backward {health-matching} method has the advantage of yielding a closed-form expression for biological age. However, when the implied biological age falls below the initial age $x_0$, its calculation requires extrapolation of the frailty curve to time $t<0$ or age $x<x_0$, where the Gompertz-Makeham law may not apply. This limitation becomes more pronounced in the vitality model, as we show in the next subsection.

\begin{figure}[ht!]
    \centering
    \includegraphics[width=\linewidth,trim={1cm 0 1cm 0},clip]{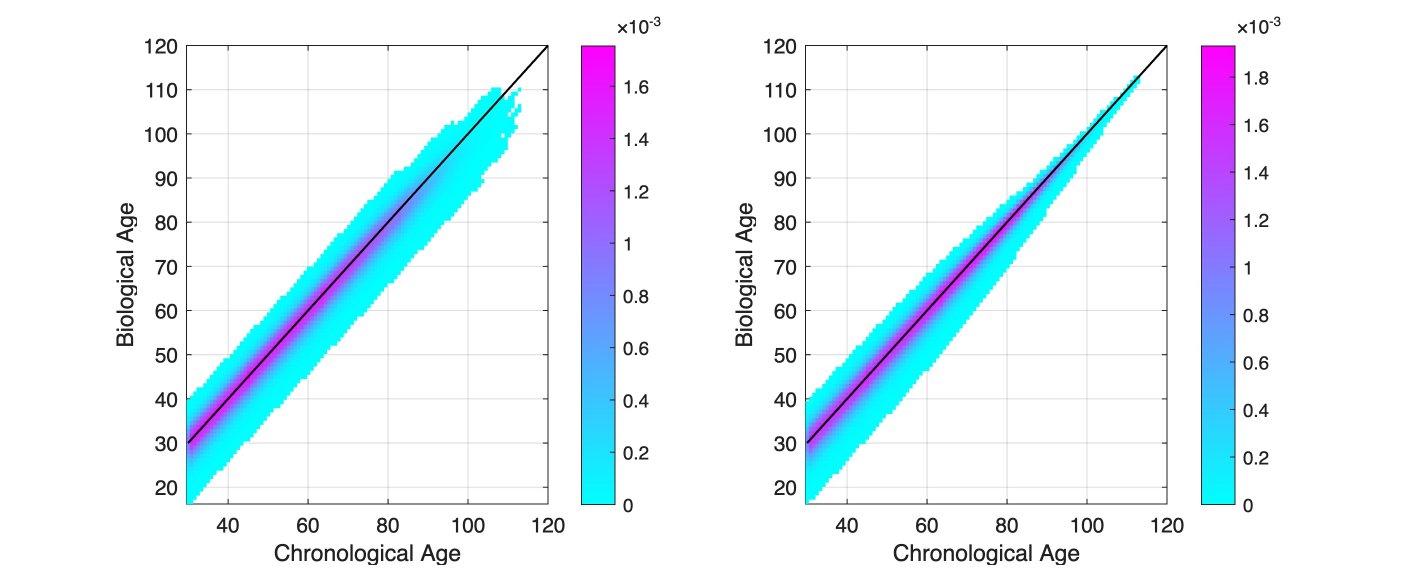}
    \caption{Simulated values of biological age against chronological age under the reliability model using the {health-matching} method (left) and the {age-shifting} method (right).}
    \label{fig:BA_heatmap_reliability}
\end{figure}

Figure~\ref{fig:BA_heatmap_reliability} shows the simulated values of biological age against chronological age under the reliability model for both methods. {In each panel, the horizontal axis is the chronological age $x_0+t$, the vertical axis is the implied biological age, and the colour indicates the density of simulated individuals, so that points on the $45^\circ$ line correspond to individuals whose biological age equals their chronological age.}

For the backward {health-matching} method, the simulated values cluster closely around the $45^\circ$ line. {Because the backward biological age depends only on an individual's current frailty, the vertical spread at each chronological age reflects the heterogeneity in frailty across the population. Individuals who are healthier than average lie below the line and are biologically younger, while frailer individuals lie above it. This spread remains roughly constant across ages, so the mapping is well behaved throughout.}

For the forward {age-shifting} method, the two ages converge along the $45^\circ$ line. {This convergence arises because the forward biological age adjusts chronological age by the difference between the population and individual remaining life expectancies, and this difference shrinks as remaining lifetimes compress at older ages, so the two ages coincide more closely at advanced ages.} Overall, both methods produce stable and interpretable mappings of biological age from chronological age for ages 35-120.

\subsection{Biological age under the vitality model} \label{sec:BioAge_Vitality}

The vitality model also provides a natural framework for defining biological age. Similar to the reliability model, we consider two methods: a backward method based on matching vitality levels, and a forward method that adjusts chronological age {using projected remaining lifetimes or death times.}

For an individual with a certain level of vitality, their backward {health-matching} biological age {$\text{BA}_{hm}$} can be determined by matching this level to the vitality trajectory of an average individual in the population. The formal mathematical definition is given as follows:
\begin{definition} \normalsize \label{def:BA_Back_Vitality}
Under the vitality model, for an individual $i$ with vitality level $V_i(t)$ at time $t$, the backward {health-matching} biological age is
\begin{align}\label{eq:gompertz_BA_vitality}
    \text{BA}_{hm}(V_i(t))
    = x_0 + \left\{ \tau : V_i(t) = V(\tau ; 0, \mathbb{E}[V_0]) \right\},
\end{align}
where $V(\tau; t, v)$ is a deterministic decreasing function of $\tau$ that determines the expected vitality level at time $\tau$ for an individual whose vitality at time $t<\tau$ is $v$.
\end{definition}

Under the vitality model, since the time of death depends only on when vitality first reaches zero, the forward method for the vitality model does not rely on the expectation of remaining lifetime, as in Definition \ref{def:BA_Forw_Reliability}. The forward {age-shifting} biological age {$\text{BA}_{as}$} is obtained by adjusting the individual's chronological age with the difference in projected passage time of zero vitality. The formal mathematical definition is given as follows:
\begin{definition} \normalsize \label{def:BA_Forw_Vitality}
Under the vitality model, for an individual $i$ with vitality level $V_i(t)$ at time $t$, the forward {age-shifting} biological age is
\begin{align}\label{eq:gompertz_BA_vitality2}
    \text{BA}_{as}(t, V_i(t)) = x_0 + t + \left( \left\{ \tau \ge 0 : V(\tau; 0,\mathbb{E}[V_0]) = 0\right\} - \left\{ \tau_i \ge 0 : V(\tau_i; t, V_i(t))  = 0 \right\}     \right),
\end{align}
where $V(\tau; t, v)$ is a deterministic decreasing function of $\tau$ that determines the expected vitality level at time $\tau$ for an individual whose vitality at time $t<\tau$ is $v$.
\end{definition}
\noindent A detailed mathematical comparison of Definitions~\ref{def:BA_Back_Vitality} and \ref{def:BA_Forw_Vitality} is provided in Appendix \ref{app:gompertz_BA}.

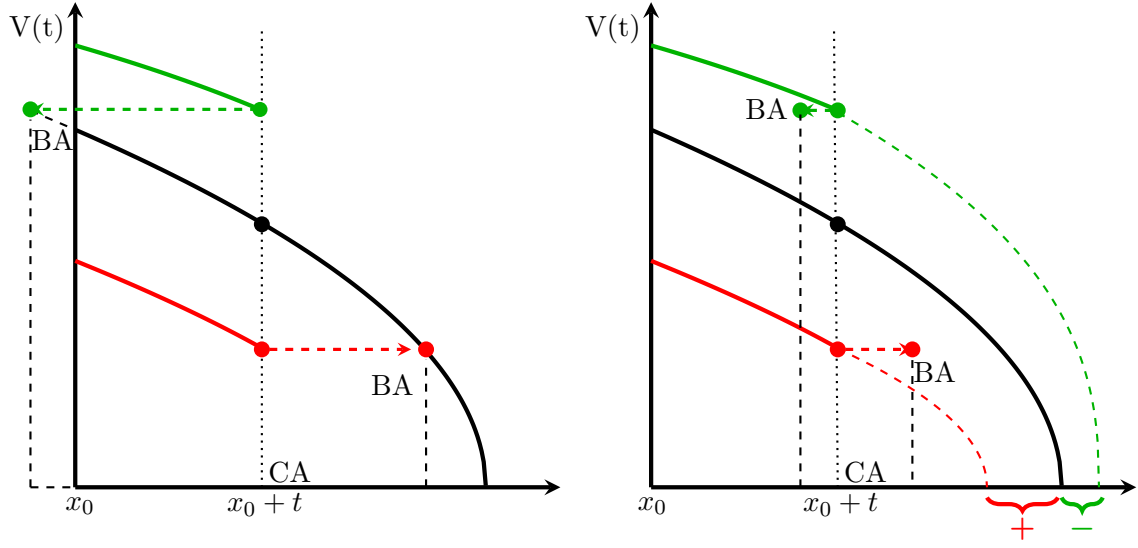
\begin{figure}[ht!]
\centering

\begin{tikzpicture}
\begin{groupplot}[
group style={
        group size=2 by 1,              % 2 columns, 1 row
        horizontal sep=1.2cm,           % gap between the two plots
        group name=plots,               % for referencing if needed
    },
    axis lines=left,
    ylabel style={at={(0.05,0.95)},rotate=90,anchor=south},
    xtick=\empty, ytick=\empty,
    xmin=0, xmax=130,
    ymin=0, ymax=95,
    width=8cm, height=8cm,
    clip=false,
    axis line style={ultra thick},
    title style={at={(0.5,1.05)},anchor=north}
]

%% ==================  Left Panel =================================
%% =================================================================
\nextgroupplot
\addplot[ultra thick, black, domain=0:110, samples=200] 
    {70 * (1 - x/110)^(0.5)};

\addplot[thick, black, dashed, domain=-12:0, samples=200] 
    {70 * (1 - x/110)^(0.5)};

\draw[dotted, thick] (axis cs:50,0) -- (axis cs:50, 90)
                     node[midway, above, sloped, black] {};

\node[above left] at (axis cs:0,85) {V(t)};
\node[below left] at (axis cs:8,0) {$x_0$};
\fill[black] (axis cs:50, 51.5) circle (3pt);

% =============== Less Healthy Person ================================
\addplot[ultra thick, red, domain=0:50, samples=300, smooth]
    {50*(1-(x+20)/110)^0.6};

\fill[red] (axis cs:50, 27) circle (3pt);
\fill[red] (axis cs:94, 27) circle (3pt);
\draw[dashed, red, very thick, -stealth] (axis cs:52, 27) -- (axis cs:90,27) node[pos=0.25, below, black] {};

%\draw[dashed, thick] (axis cs:72,0) -- (axis cs:72, {50*(1-72/110)^0.7-3}) node[midway, above, sloped, black] {};
\draw[dashed, thick] (axis cs:94,0) -- (axis cs:94, 25) node[midway, above, sloped, black] {};

\node at (axis cs:85, 20)  {BA};

\node[below right] at (axis cs: 49,7) {CA};
\node[below right] at (axis cs: 38,0.8) {$x_0+t$};

% ================ Healthy Person ======================================
\addplot[ultra thick, green!70!black, domain=0:50, samples=300, smooth]
    {85*(1-x/120)^0.3  + 1.5}
    node[pos=0.99, circle, fill=green!70!black, inner sep=2pt] {};
\draw[dashed, thick] (axis cs:-12,0) -- (axis cs:-12, 72)
                     node[midway, above, sloped, black] {};
\draw[dashed, thick] (axis cs:-12,0) -- (axis cs:0, 0)
                     node[midway, above, sloped, black] {};
%\draw[dashed, thick] (axis cs:38,0) -- (axis cs:38, 72) node[midway, above, sloped, black] {};
\fill[green!70!black] (axis cs:-12,74) circle (3pt);
\draw[dashed, green!70!black, very thick, -stealth] (axis cs:49,74) -- (axis cs:-12,74) node[pos=0.25, below, black] {};
%\node at (axis cs:60, 74)  {CA};
\node at (axis cs:-6, 68)  {BA};

%% ================ Right Penal ======================================

\nextgroupplot
\addplot[ultra thick, black, domain=0:110, samples=200] {70 * (1 - x/110)^(0.5)};
\node[above left] at (axis cs:0,85) {V(t)};
\node[below left] at (axis cs:7,0) {$x_0$};
\fill[black] (axis cs:50, 51.5) circle (3pt);

\draw[dotted, thick] (axis cs:50,0) -- (axis cs:49, 90)
                     node[midway, above, sloped, black] {};
                     
%% =========== Less Healthy Person =======================================
\addplot[ultra thick, red, domain=0:50, samples=300, smooth]
    {50*(1-(x+20)/110)^0.6};

\fill[red] (axis cs:50, 27) circle (3pt);
\fill[red] (axis cs:70,27) circle (3pt); 

\addplot[thick, red, dashed, domain=50:90, samples=200] 
    {45 * (1 - (x+20)/110)^(0.5) };

\draw[ultra thick,decorate,red, decoration={brace,amplitude=5pt,raise=0pt,mirror}]
    (axis cs:90,-2) -- node[below=3pt] {$\boldsymbol{+}$} (axis cs:109,-2);
\draw[dashed, red, very thick, -stealth] (axis cs:52,27) -- (axis cs:70,27) node[pos=0.25, below, black] {};

%\draw[ultra thick,decorate,red, decoration={brace,amplitude=5pt,raise=2pt,mirror}]
%    (axis cs:74,21) -- node[below=8pt] {} (axis cs:92,21);

\draw[dashed, thick] (axis cs:70,1.5) -- (axis cs:70, 25)
                     node[midway, above, sloped, black] {};

\node at (axis cs:76, 22.5)  {BA};

\node[below right] at (axis cs: 49,7) {CA};
\node[below right] at (axis cs: 38,0.8) {$x_0+t$};

%% ============ Healthy Person ===========================================
\addplot[ultra thick, green!70!black, domain=0:50, samples=300, smooth]
    {85*(1-x/120)^0.3  + 1.5};

\addplot[thick, green!70!black, dashed, domain=50:120, samples=200] 
    {91*(1-x/120)^0.4  };

\draw[ultra thick,decorate,green!70!black, decoration={brace,amplitude=5pt,raise=0pt,mirror}]
    (axis cs:110,-2) -- node[below=3pt] {$\boldsymbol{-}$} (axis cs:121,-2);

\fill[green!70!black] (axis cs:40,73.8) circle (3pt);
\fill[green!70!black] (axis cs:50,73.8) circle (3pt);

\draw[dashed, green!70!black, very thick,  stealth-] (axis cs: 40,73.8 ) -- (axis cs:50,73.8) node[pos=0.25, below, black] {};

\draw[dashed, thick] (axis cs:40,0) -- (axis cs:40, 73)
                     node[midway, above, sloped, black] {};

\node at (axis cs:31, 74)  {BA};

\end{groupplot}
\end{tikzpicture}
\caption{Visual illustration of how biological age is determined under the vitality model using the backward {health-matching} method (left) and the forward {age-shifting} method (right). {The solid black curve is the average vitality trajectory and the green and red curves are a healthier and a frailer individual, respectively, with filled circles marking chronological age (CA) and biological age (BA). In the right panel, the dashed segments show the projected future vitality, and the $+$ and $-$ signs indicate whether the adjustment is added to or subtracted from CA to obtain BA.}}
\label{fig:BA_illustration_vitality}
\end{figure}

Figure \ref{fig:BA_illustration_vitality} provides a graphical illustration of Definitions \ref{def:BA_Back_Vitality} and \ref{def:BA_Forw_Vitality}. As in the reliability model, the backward method requires extrapolation of the vitality curve when the implied biological age falls below the initial age $x_0$. When the depletion rate specified by the Gompertz law, the vitality level changes slowly for $t<0$, which leads to an undefined biological age when $V_i(t) > \mathbb{E}[V_0] + {b}/{\ln c}$ under the backward method. The forward method avoids this issue, but it also has the limitation that biological age cannot exceed the expected remaining lifetime of the average individual at the same chronological age.

\begin{figure}[ht!]
    \centering
    \includegraphics[width=\linewidth,trim={1cm 0 1cm 0},clip]{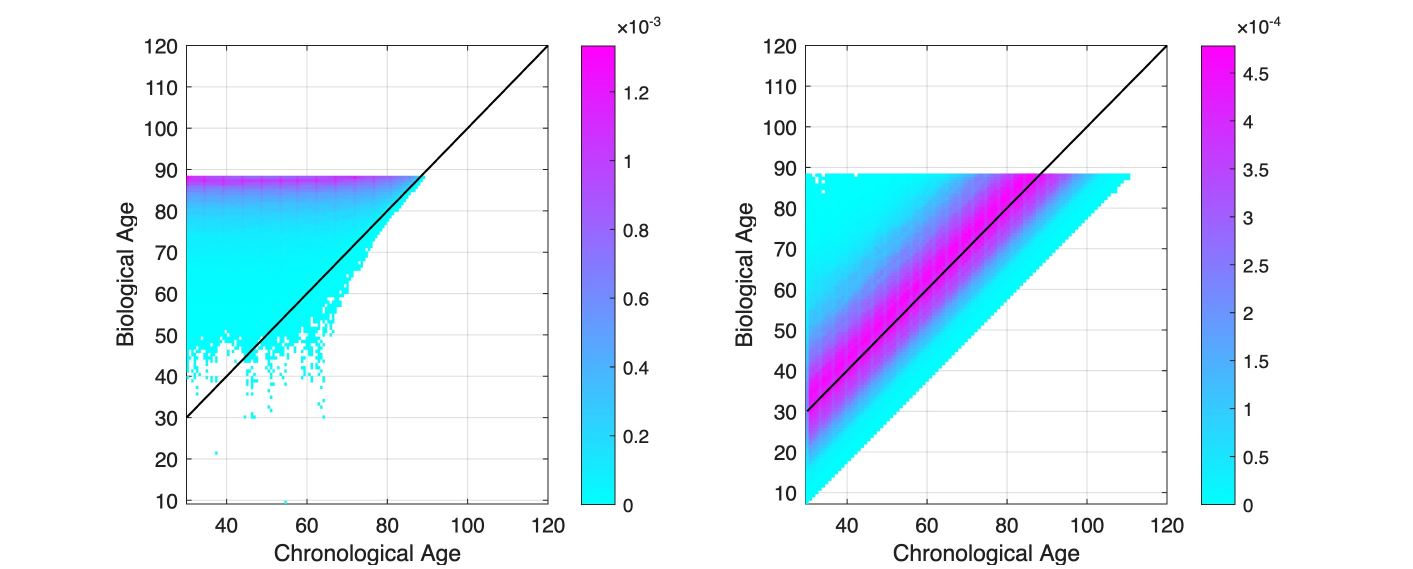}
    \caption{Simulated values of biological age against chronological age under the vitality model using the {health-matching} method (left) and the {age-shifting} method (right).}
    \label{fig:BA_heatmap_vitality}
\end{figure}

Figure~\ref{fig:BA_heatmap_vitality} presents the simulated values of biological age against chronological age under the vitality model for both methods. For both methods, biological age increases with chronological age but cannot exceed the value implied by the expected remaining lifetime of an average individual. This constraint creates a visible upper bound in the heatmap shown, and makes the mapping from chronological age to biological age unreliable under the vitality model.

{
\subsection{Comparison of biological age constructions}\label{sec:BioAge_Compare}

The health-matching and age-shifting methods can assign different biological ages to the same individual, which raises the question of how the two constructions relate and when they agree. The health-matching method has the interpretation most common in the literature, where the biological age is the chronological age of a reference individual with the same health level as the individual of interest. The age-shifting method also admits a clear interpretation. For example, a heavy smoker aged $50$ with a remaining life expectancy of $20$ years, compared with an average $50$-year-old whose remaining life expectancy is $27$ years, has a seven-year shortfall that translates into a biological age of $50+7=57$.

A third construction matches the individual on remaining lifetime rather than on current health. Following \cite{sluiskes2024accelerage}, the biological age is defined as the chronological age at which an average person has the same remaining lifetime as the individual, which we call the \emph{lifetime-matching} method:
\begin{definition} \normalsize \label{def:BA_lm_reliability}
Under the reliability model, for an individual $i$ with frailty level $F_i(t)$ at time $t$, the lifetime-matching biological age is
\begin{align}\label{eq:gompertz_BA_reliability3}
    \text{BA}_{lm}(t,F_i(t)) = x_0 + \left\{t_i: \mathbb{E}[T_{x_0+t_i}(\mathbb{E}[F(t_i;0,\mathbb{E}[F_0])])] = \mathbb{E}[T_{x_0+t}(F_i(t))] \right\},
\end{align}
where $T_x(F)$ is the remaining lifetime of an individual aged $x$ with frailty level $F$.
\end{definition}
\begin{definition} \normalsize \label{def:BA_lm_vitality}
Under the vitality model, for an individual $i$ with vitality level $V_i(t)$ at time $t$, the lifetime-matching biological age is
\begin{align}\label{eq:gompertz_BA_vitality3}
    \text{BA}_{lm}(t,V_i(t)) = x_0 + \left\{t_i: \mathbb{E}[T_{x_0+t_i}(\mathbb{E}[V(t_i)])] = \mathbb{E}[T_{x_0+t}(V_i(t))] \right\},
\end{align}
where $T_x(v)$ is the remaining lifetime of an individual currently aged $x$ with vitality level $v$.
\end{definition}
\noindent The lifetime-matching method combines features of the other two constructions: it is based on the projected future lifetime, as in the age-shifting method, and is obtained by matching to a reference individual, as in the health-matching method. Which of the three constructions coincide depends on the model.

We formalise these relationships in the two corollaries below:
\begin{corollary} \normalsize \label{cor:Gompertz_newBA_reliability}
Under the reliability model, the lifetime-matching biological age $\text{BA}_{lm}$ is equivalent to the health-matching biological age $\text{BA}_{hm}$, and both differ from the age-shifting biological age $\text{BA}_{as}$.
\end{corollary}
\noindent Intuitively, the remaining lifetime in the reliability model is a monotone function of the current frailty, so matching on health and matching on lifetime yield the same reference age. The proof is given in Appendix \ref{app:gompertz_analysis}.
\begin{corollary} \normalsize \label{cor:Gompertz_newBA_vitality}
Under the vitality model, the lifetime-matching biological age $\text{BA}_{lm}$ is equivalent to the age-shifting biological age $\text{BA}_{as}$ if depletion is deterministic, and all three constructions coincide ($\text{BA}_{hm}=\text{BA}_{as}=\text{BA}_{lm}$) if vitality follows a random walk with negative constant drift.
\end{corollary}
\noindent By contrast, the remaining lifetime in the vitality model depends on both current vitality and age, so the health-matching method separates from the other two under the Gompertz depletion of the baseline specification. The proof is given in Appendix \ref{app:gompertz_analysis}.

The differences are most pronounced for the vitality model, as shown in Figure~\ref{fig:BA_heatmap_vitality}. Under that specification, the health-matching and age-shifting methods disagree, and the health-matching method is undefined for individuals whose vitality far exceeds the average initial level. In addition, because the constructions are referenced to the average trajectory, the vitality biological age cannot exceed $x_0 + \mathbb{E}[T_{x_0}(\mathbb{E}[V_0])]$, the age at which the average person would die.

As illustrated in \cite{zhu2025mortality}, other specifications of the vitality model can also produce the Gompertz law. Under an age-homogeneous specification, the initial vitality follows a Gompertz distribution, $\mathbb{P}(V_0<v) = 1- e^{-\eta(e^{v} -1)}$, and vitality depletes linearly, $\mu_{x_0}(t)=\delta$. The remaining lifetime is then $v/\delta$, which is independent of age, and the three constructions coincide according to Corollary \ref{cor:Gompertz_newBA_vitality}. However, the upper bound is intrinsic to referencing an average individual of finite lifespan, and remains a genuine limitation of biological age under the vitality model that the reliability model does not share.
}

\subsection{Subjective survival beliefs}\label{sec:BioAge_Subjective}

A well-established empirical finding in the literature is that individuals tend to underestimate survival at younger ages and overestimate it at older ages \citep{ludwig2013parsimonious, groneck2016life, kalwij2021accuracy, apicella2024behavioral}. As demonstrated in \cite{zhu2025mortality}, the vitality-based approach can account for these subjective survival biases. More specifically, these biases will emerge when an individual underestimates both their initial vitality and their depletion rate. Similarly, the reliability-based approach can capture the same phenomenon if an individual overestimates their initial frailty $F_0$ and underestimates the subsystem failure rate $r$. {Under both approaches, the individual forms their survival belief at the initial age $x_0$, so that the initial vitality $V_0$ and the initial frailty $F_0$ represent their vitality and frailty, respectively, at the moment of self-assessment.}

Another observation from \cite{zhu2025mortality} regarding the vitality-based approach is the disparity between subjective beliefs about health status and life expectancy. The paper shows that under a Gompertz depletion rate,
\begin{align*}
    \mathbb{E}[T_{x_0}(V_0)] \leq T_{x_0}(\mathbb{E}[V_0]),
\end{align*}
which implies that the average life expectancy is \emph{smaller} than the life expectancy of an individual with an average vitality level. Under the reliability-based approach, we found a similar distinction but in the opposite direction, as stated in the following proposition:
\begin{prop}\label{prop:gompertz_expectancy}
    Let $F_0 \sim \mathbb{F}$, where $\mathbb{F}$ is a distribution function with positive support, and let $T_{x_0}(F_0)$ denote the remaining lifetime of a person aged $x_0$ with an initial number of dysfunctional subsystems~$F_0$. If $F(t)$ evolves according to \eqref{eq:gompertz_reliability_mu}, then
    \begin{align*}
        \mathbb{E}[T_{x_0}(F_0)] \geq \mathbb{E}[T_{x_0}(\mathbb{E}[F_0])].
    \end{align*}
\end{prop}
Proposition \ref{prop:gompertz_expectancy} therefore shows that the average life expectancy is \emph{larger} than the life expectancy of an individual with an average initial frailty level. The proof is given in Appendix \ref{app:gompertz_analysis}. Although the reliability-based approach yields a conclusion opposite to that of the vitality-based approach, both indicate that relying on average life expectancy can be misleading, and that other remaining lifetime measures, such as quantile-based ones, may be more appropriate.

%%%%%%%%%%%%%%%%%%%%%%
%%% Section 5: Conclusion
%%%%%%%%%%%%%%%%%%%%%%

\section{Conclusion} \label{sec:Conclusion}

This paper presented a unified actuarial view of the vitality-based and reliability-based approaches for mortality modelling. We showed that, despite originating from different biological mechanisms, both approaches naturally generate Gompertz mortality under suitable assumptions and can be extended to reproduce the Makeham law. Using Canadian mortality data, we calibrated both approaches and compared their empirical behaviour across several specifications. We also developed parallel definitions of biological age and examined subjective survival beliefs under each approach. Our analyses highlighted both the conceptual similarities and the fundamental differences between the two approaches.

The vitality-based approach nevertheless has limitations. The deterministic depletion function $\mu_{x_0}(t)$ provides a reasonable population-level description but is difficult to justify at the individual level, where vitality trajectories can be structurally changing and genetically determined. Future work may explore alternative depletion functions that allow non-monotonic patterns or link closely to empirical biometrics, such as genetic markers, cognitive measures, or gender-related characteristics. Another direction is to reconsider the role of stochastic diffusion, including the use of switching regimes to capture changes in mortality dynamics.

The reliability-based approach also has limitations. The assumption that subsystem failures evolve independently may be unrealistic. Network-based frailty models, such as those proposed by \cite{barabasi2011network}, \cite{mitnitski2017aging} and \cite{rutenberg2018unifying}, offer a promising direction for directly capturing interactions among subsystems and studying more complex patterns of health deterioration. Another possible extension is to introduce a time-dependent form for $\mu_x(t)$, such as $\logit \mu_x(t) = \ln \frac{F_0}{N-F_0} + c\, t$, where $c$ is a constant and $R_0 := \ln \frac{F_0}{N-F_0} \in \mathbb{R}$ represents the distribution of initial health states.

A common limitation of both approaches is the reliance on a single mathematical process to represent ageing. Future research could aim to link vitality or frailty to empirical measures of intrinsic capacity, such as grip strength, lung function, haemoglobin levels, endocrine markers, and cardiovascular indicators, as documented in \citep{beard2019structure, beard2022intrinsic, si2023life, chen2025intrinsic, numbers2025intrinsic}. In addition, the definitions of biological age and subjective survival beliefs discussed in this paper can be further incorporated into actuarial decision processes, including annuity demand, insurance underwriting, retirement planning, and longevity risk management.

There is no single definitive framework that can fully explain human ageing and mortality. Mechanistic ageing models nonetheless offer a natural and intuitive way to describe the biological foundations of mortality dynamics. This paper provides a basis for incorporating vitality-based and reliability-based modelling ideas into actuarial science, and we hope it will support further exploration of mechanistic perspectives in future research.

\bibliography{reference}

\begin{appendices}

\section{} \label{app:gompertz_reliability}

Denote $m(t;F_0,N) = \mathbb{E}[F(t;0, F_0, N)|F_0]$ and $\nu(t;F_0,N) = \text{Var}(F(t;0,F_0, N)|F_0)$ as the conditional expectation and conditional variance of the number of dysfunctional subsystems at time $t$, given an initial $F_0$ at $t=0$. The parameter $N$ is included explicitly to emphasize that these moments depend on the total number of subsystems; in later derivations, we will often suppress both $F_0$ and $N$ for notational brevity. For large $N$, \cite{nielsen2024gompertz} employ the linear noise approximation to derive the following approximations:
\begin{align}\label{eq:gompertz_approx}
    m(t) &\approx \frac{N \cdot F_0}{ F_0 + (N-F_0)e^{-rNt} }, \qquad  \nu(t) \approx F_0 \cdot e^{2rNt}\cdot  \left(1 - e^{-rNt}\right).
\end{align}
Here we re-derive these approximations directly from Kolmogorov's forward equation.

\subsection{Dynamics of $m(t)$ and $\nu(t)$} \label{app:gompertz_reliability_ODE}

By Kolmogorov's forward equation, we have
\begin{align*}
    \frac{\diff}{\diff t}{}_tp_x^{i,k} = r\cdot (k-1)\cdot (N-k+1)\cdot {}_{t}p_{x}^{i, k-1} - r\cdot k\cdot (N-k)\cdot {}_tp_{x}^{i, k}
\end{align*}
for $i,k \in [0,N]$ with ${}_tp_x^{i, k} = 0$ if $k<F_0$. By the definition of expectation and variance:
\begin{align*}
    m(t;F_0, N) = \sum_{k=0}^{N} k\cdot {}_tp_{x_0}^{F_0, k}, \qquad \nu(t;F_0, N) = \left(\sum_{k=0}^{N} k^2\cdot {}_tp_{x_0}^{F_0, k}\right) - \left(\sum_{k=0}^{N} k\cdot {}_tp_{x_0}^{F_0, k}\right)^2.
\end{align*}
The ordinary differential equation (ODE) of $m(t)$ can be derived as:
\begin{align*}
    \frac{\diff}{\diff t}m(t) &= \sum_{k=0}^{N} k\cdot \left(\frac{\diff}{\diff t}{}_tp_{x_0}^{F_0, k}\right)= \sum_{k=1}^{N} k\cdot \left[ r\cdot (k-1)\cdot (N-k+1)\cdot {}_tp_{x_0}^{F_0, k-1} - r\cdot k\cdot (N-k)\cdot {}_tp_{x_0}^{F_0, k}    \right]\\
    &= \left( \sum_{k=0}^{N-1} (k+1)\cdot r\cdot k\cdot (N-k)\cdot {}_tp_{x}^{F_0, k}\right) - \left( \sum_{k=1}^{N} k \cdot r\cdot k\cdot (N-k)\cdot {}_tp_{x}^{F_0, k}  \right)\\
    &= \sum_{k=0}^{N} r\cdot k\cdot (N-k) \cdot {}_tp_{x_0}^{F_0, k}\\
    &= r\cdot m(t)\cdot (N-m(t)) - r\cdot \nu(t)
\end{align*}
and similarly for $\nu(t)$:
\begin{align*}
    \frac{\diff}{\diff t}\nu(t) &= \left(\sum_{k=0}^{N} k^2\cdot \left(\frac{\diff}{\diff t}{}_tp_{x_0}^{F_0, k}\right)\right) - \frac{\diff}{\diff t}\left(m(t)^2  \right)\\
    &=  \left(\sum_{k=1}^{N} k^2\cdot \left[ r\cdot (k-1)\cdot (N-k+1)\cdot {}_tp_{x_0}^{F_0, k-1} - r\cdot k\cdot (N-k)\cdot {}_tp_{x_0}^{F_0, k}    \right]\right) - 2\cdot m(t) \cdot \frac{\diff}{\diff t}m(t)\\
    &= \left(\sum_{k=0}^{N} (2k+1) \cdot r\cdot k \cdot (N-k)\cdot {}_tp_{x_0}^{F_0, k}\right) - 2\cdot m(t) \cdot \left[ r\cdot m(t) \cdot (N-m(t)) - r\cdot \nu(t)  \right]\\
    &= r\cdot m(t)\cdot (N-m(t)) + r\cdot [ 2N-1 +2m(t) ]\cdot \nu(t) + 2r\cdot m(t)^3 - 2r\cdot \mathbb{E}[F(t)^3|F_0]
\end{align*}
Therefore, the ODE for the $i$-th conditional moment of $F(t)$ depends on the $(i+1)$-th moment. Thus, the system does not admit a closed-form solution.

\subsection{Approximation of $m(t)$ and $\nu(t)$} \label{app:gompertz_reliability_approx}

Since both $m(t)$ and $\nu(t)$ become infinite as $N\rightarrow \infty$, it is more convenient to work with the rescaled quantities, $m(t)/N$ and $\nu(t)/N$. To ensure that the dynamics of $m(t)/N$ are well defined in the limit $N\rightarrow \infty$, we define the rescaled growth rate $c = rN$ and assume that the value of $c$  is independent of $N$. 

By the Law of large numbers (e.g., Chapter 11, Theorem 2.1 of \cite{ethier2009markov}), for every fixed $t\geq 0$, and any fixed initial fraction $\phi_0 = F_0/N$,
\begin{align*}
    \lim_{N\rightarrow \infty} \sup_{s\leq t} \left| \frac{F(s;0, F_0, N)}{N} - \phi(s)\right| = 0
\end{align*}
where $\phi(t)$ solves the deterministic ODE
\begin{align*}
    \frac{\diff}{\diff t}\phi(t) = c\cdot \phi(t)\cdot (1-\phi(t)), \quad \phi(0) = \phi_0.
\end{align*}
The explicit solution is
\begin{align*}
    \phi(t) = \frac{\phi_0}{\phi_0 + (1-\phi_0)e^{-ct}}.
\end{align*}
Thus, for large $N$,
\begin{align*}
    m(t) \approx \frac{NF_0}{F_0 + (N-F_0)e^{-rNt}}.
\end{align*}
For small $t$ and $F_0 \ll N$, then $F_0 + (N-F_0)e^{-rNt}\approx Ne^{-rNt}$ and therefore $m(t) \approx F_0e^{rNt}$.

For approximate the variance, consider the ODE for $\nu(t)$ when $N$ is large and $F(t)\ll N$. In this case, higher-order conditional moments of $F(t)$ may be neglected, and $N-m(t) \approx N$. Then, substituting the approximation of $m(t) \approx F_0e^{rNt}$ yields
\begin{align*}
    \frac{\diff}{\diff t}\nu(t) &\approx r \cdot m(t) \cdot N + 2r\cdot N \cdot\nu(t) \\
    \implies \nu(t) &\approx F_0\cdot e^{2rNt}\cdot \left(1 - e^{-rNt}\right)
\end{align*}
which exactly matches the approximations reported in \cite{nielsen2024gompertz}.

\subsection{A Multi-State Model with Mortality} \label{app:gompertz_reliability_multistate}

Without loss of generality, we define the number of dysfunctional subsystems at the moment of death to be equal to the number present immediately prior to death. Let $\mathcal{S}(t)$ denote the state at time $t$, with $\mathcal{S}(t)\leq N$ indicating that the individual is alive and $\mathcal{S}(t)>N$ indicating death (absorbing states). The multi-state model is graphically illustrated below: 
\begin{figure}[H]
\begin{center}
    	\begin{tikzpicture}[
			rounded corners=5pt,
			inner sep=7pt,
			node distance=0.7cm]
			\tikzstyle{my node}=[draw,minimum height=1cm,minimum width=3cm, text width=2.5cm]
			\node[my node] (f0){$\mathcal{S}(t) = 0$ $ (F(t) = 0)$};
			\node[my node,right=of f0](f1){$\mathcal{S}(t) = 1$ $(F(t) = 1)$};
            \node[draw=none, right=of f1] (fk) {$\cdots \cdots$};
            \node[my node,right=of fk](fn){$\mathcal{S}(t) = N$ $(F(t) = N)$};
            \node[my node, below=of f0](d0) {Death $\qquad \mathcal{S}(t) = N+1$};
            \node[my node, below=of f1](d1) {Death $\qquad \mathcal{S}(t) = N+2$};
            \node[my node, below=of fn](dn) {Death $\qquad \mathcal{S}(t) = 2N+1$};
            \node[draw=none, below= 1.5cm of fk](dk) {$\cdots \cdots$};
			\draw[->] (f0) -- (d0);
			\draw[->] (f1) -- (d1);
            \draw[->] (fn) -- (dn);
            \draw[->] (f0) -- (f1);
            \draw[->] (f1) -- (fk);
            \draw[->] (fk) -- (fn);
		\end{tikzpicture}
\end{center}
    \caption{A multi-state structure with death states for the subsystem failure process.}
    \label{fig:gompertz_model_mortality}
\end{figure}
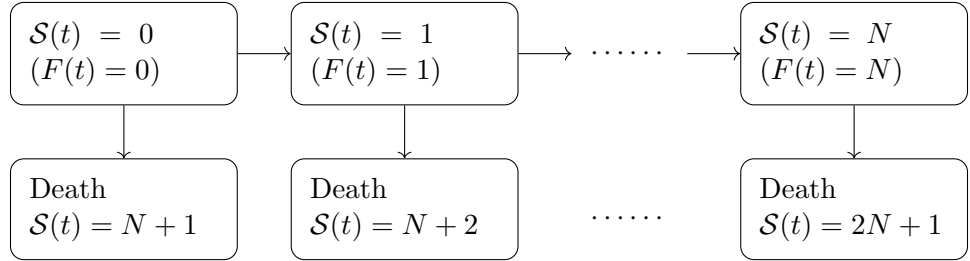
Again, for notational convenience in the derivations, we all $\mathcal{S}(t)$ to take values in $\{0, 1, \cdots, 2N+1\}$, while imposing zero probability of transitioning to any state below the initial value $F_0$. Define the conditional expectation and variance of the number of dysfunctional subsystems among individuals who are still alive at time $t$ as
\begin{align*}
m(t; F_0, N) &= \mathbb{E}[F(t) \times \mathbb{I}(\mathcal{S}(t)\leq N)|F(0) = F_0]\\
\nu(t; F_0, N) &= \text{Var}\left( F(t)\times \mathbb{I}(\mathcal{S}(t)\leq N)|F_0 \right)
\end{align*}
The indicator function is essential because the force of mortality is defined only on the surviving population. Assume that the force of mortality at time $t$ is
\begin{align*}
    \mu(t) = \kappa \cdot \frac{F(t)}{N}.
\end{align*}
Applying Kolmogorov's forward equation to the transition probabilities, we have
\begin{align*}
    \frac{\diff }{\diff t}m(t) &= \sum_{k=0}^N k\times \frac{\diff}{\diff t}{}_tp_{x_0}^{F_0,k}\\
    &=\sum_{k=0}^{N} k\times\left[  r(k-1)(N-k+1){}_tp_{x_0}^{F_0, k-1} - rk(N-k){}_tp_{x_0}^{F_0, k} - \kappa \cdot \frac{k}{N} {}_tp_{x_0}^{F_0, k}\right]\\
    &= r\cdot m(t)\cdot (N-m(t)) - r\cdot \nu(t) - \kappa \left(\frac{\nu(t)}{N} + \frac{m(t)^2}{N}\right)
\end{align*}
For large $N$, the additional term introduced by the mortality can be neglected, and the approximations developed in the previous section remain applicable here.

\section{Biological ages under the vitality model}\label{app:gompertz_BA}
The forward {age-shifting} biological age is given in Definition \ref{def:BA_Forw_Vitality} as
\begin{align*}
        \text{BA}_{as}(t, V_i(t)) = x_0 + t + \left(\left\{\tau\geq 0: V(\tau;0,\mathbb{E}[V_0]) = 0\right\} - \left\{\tau_i\geq 0: V(\tau_i; t, V_i(t))  = 0 \right\}     \right).
\end{align*}
Assuming Gompertz decay $\mu(t) = bc^t$, we have
\begin{align*}
    &\mathbb{E}[V_0] - \frac{b}{\ln c}\left(c^{\tau}-1\right) = 0 \implies \tau = \frac{1}{\ln c} \ln \left(\frac{\mathbb{E}[V_0] \cdot \ln c}{b} + 1\right)\\
    &V_i(t) - \frac{b}{\ln c}\left(c^{\tau_i} - c^{t}\right) = 0 \implies \tau_i = \frac{1}{\ln c} \ln \left(\frac{V_i(t)\cdot \ln c}{b} + c^t\right).
\end{align*}
Thus,
\begin{align*}
     \text{BA}_{as}(t, V_i(t)) &= x_0 + t + \frac{1}{\ln c} \ln \left(\frac{\mathbb{E}[V_0] \cdot \ln c}{b} + 1\right) - \frac{1}{\ln c} \ln \left(\frac{V_i(t)\cdot \ln c}{b} + c^t\right)\\
     &=x_0 + t + \frac{1}{\ln c} \ln\left( \ln c\cdot \frac{\mathbb{E}[V_0] - V_i(t)}{b}\frac{b}{V_i(t)\ln c + bc^t} + \frac{b-bc^t}{V_i(t)\ln c + bc^t} + 1    \right) .
\end{align*}
This expression is well-defined whenever $V_i(t)>0$ and $c>1$.\\

On the other hand, the backward {health-matching} method defines the biological age as
\begin{align*}
    \text{BA}_{hm}(V_i(t)) = x_0 + \left\{\tau: V(\tau;0, \mathbb{E}[V_0]) = V_i(t)\right\}
\end{align*}
such that
\begin{align*}
    \mathbb{E}[V_0] - \frac{b}{\ln c}\left(c^{\tau}-1\right) = V_i(t) \implies \tau = \frac{1}{\ln c} \ln \left(\ln c\cdot \frac{\mathbb{E}[V_0] -  V_i(t)}{b} + 1\right).
\end{align*}
Thus,
\begin{align*}
    \text{BA}_{hm}(V_i(t)) = x_0 +\frac{1}{\ln c} \ln \left(\ln c\cdot \frac{\mathbb{E}[V_0] -  V_i(t)}{b} + 1\right).
\end{align*}
The solution of $\text{BA}_{hm}(V_i(t))$ can be a complex number if
\begin{align*}
    \ln c\cdot \frac{\mathbb{E}[V_0] -  V_i(t)}{b} + 1 < 0 \quad \text{or} \quad V_i(t) >  \mathbb{E}[V_0] + \frac{b}{\ln c} .
\end{align*}
We therefore can conclude that the two definitions of biological age under the vitality model are not equivalent. For instance, at $t=0$, as long as $\mathbb{E}[V_0] \neq V_i(0) >0$, we get
\begin{align*}
    \text{BA}_{as}(0,V_i(0)) &= x_0  + \frac{1}{\ln c}\ln \left( \ln c\frac{\mathbb{E}[V_0] - V_i(0)}{V_i(0)\ln c + b} + 1   \right) \\
    &\neq x_0  + \frac{1}{\ln c}\ln \left( \ln c\frac{\mathbb{E}[V_0] - V_i(0)}{b} + 1   \right) = \text{BA}_{hm}(V_i(0)).
\end{align*}

{
\section{Proofs}\label{app:gompertz_analysis}
\subsection{Proof of Corollary \ref{cor:Gompertz_newBA_reliability}}
Under the reliability model, the remaining lifetime depends only on the current frailty, not on age, i.e., $T_x(F) \overset{\text{d}}{=} T_y(F)$ for any $x$ and $y$. Moreover, because the only randomness in the frailty level enters through $F_0$, we have $\mathbb{E}[F(t;0,\mathbb{E}[F_0])] = F(t;0,\mathbb{E}[F_0])$. Therefore
\begin{align*}
    \text{BA}_{lm}(t,F_i(t)) &= x_0 + \left\{t_i: \mathbb{E}[T_{x_0+t_i}(\mathbb{E}[F(t_i;0,\mathbb{E}[F_0])])] = \mathbb{E}[T_{x_0+t}(F_i(t))] \right\}\\
    &= x_0 + \left\{t_i: \mathbb{E}[T_{x_0}(\mathbb{E}[F(t_i;0,\mathbb{E}[F_0])])] = \mathbb{E}[T_{x_0}(F_i(t))] \right\}\\
    &= x_0 + \left\{t_i: F(t_i;0,\mathbb{E}[F_0]) = F_i(t) \right\},
\end{align*}
which is precisely the health-matching biological age. The age-shifting biological age coincides with the health-matching one if and only if
\begin{align*}
    \mathbb{E}[T_{x_0+t-\Delta}(F(t-\Delta;0,\mathbb{E}[F_0]))] = \mathbb{E}[T_{x_0+t}(F(t;0,\mathbb{E}[F_0]))] + \Delta
\end{align*}
for all $t$ and $\Delta$, which would require a person's remaining life expectancy to decrease by exactly one year as they age by one year. This never holds under a realistic mortality law, so the age-shifting method differs from the other two in general.

\subsection{Proof of Corollary \ref{cor:Gompertz_newBA_vitality}}
We first consider deterministic depletion. Denote $\tau_i = T_{x_0+t}(V_i(t)) + t$, so that $x_0 + \tau_i$ is the age at death of the individual currently aged $x_0+t$ with vitality $V_i(t)$. Then, we have
\begin{align*}
    \text{BA}_{lm}(t,V_i(t)) &= x_0 + \left\{t_i: T_{x_0+t_i}(V(t_i;0,\mathbb{E}[V_0])) = T_{x_0+t}(V_i(t)) \right\}\\
    &= x_0 + \left\{t_i: V(t_i+\tau_i-t;0,\mathbb{E}[V_0]) = 0 \right\}\\
    &= x_0 + \left\{u_i: V(u_i;0,\mathbb{E}[V_0]) = 0 \right\} - (\tau_i - t)\\
    &= \text{BA}_{as}(t,V_i(t)),
\end{align*}
where the second line uses the definition of the death time as the time at which vitality reaches zero.

We next consider a constant depletion rate. When depletion is linear in time, so that the cumulative depletion by time $t$ is $\delta t$ ($\delta>0$), then even with a Brownian component we have $\mathbb{E}[T_{x_0+t}(v)] = v/\delta$, independent of chronological age. Therefore, we have
\begin{align*}
    \text{BA}_{lm}(t,V_i(t)) &= x_0 + \left\{t_i: \mathbb{E}[T_{x_0+t_i}(V(t_i;0,\mathbb{E}[V_0]))] = \mathbb{E}[T_{x_0+t}(V_i(t))] \right\}\\
    &= x_0 + \left\{\tau: \mathbb{E}[V(\tau;0,\mathbb{E}[V_0])] = V_i(t) \right\}\\
    &= \text{BA}_{hm}(V_i(t)),
\end{align*}
where $V(\tau;0,\mathbb{E}[V_0])$ is replaced by its expectation in the definition of $\text{BA}_{hm}$, since the diffusion in the depletion is ignored in the baseline. Also notice that
\begin{align*}
    \left\{\tau: \mathbb{E}[V(\tau;0,\mathbb{E}[V_0])] = V_i(t) \right\}  &= \frac{\mathbb{E}[V_0]}{\delta} - \frac{V_i(t)}{\delta}\\
    &= \left\{\tau\geq 0: \mathbb{E}[V(\tau; 0, \mathbb{E}[V_0])] = 0\right\} - \big(\left\{\tau_i\geq 0: \mathbb{E}[V(\tau_i; t, V_i(t))] = 0\right\}-t\big),
\end{align*}
then immediately see that $\text{BA}_{lm}$ is equivalent to the definition of $\text{BA}_{as}$ with $V$ replaced by its expectation, since the diffusion in the depletion is ignored in the baseline. Therefore, all three biological age constructions coincide.
}

\subsection{Proof of Proposition \ref{prop:gompertz_expectancy}}
From the large-$N$ approximation of Eq.\eqref{eq:F_approx}, $F(t; F_0) = \frac{NF_0}{F_0 + (N-F_0)e^{-rNt}}$, we compute the first and second partial derivatives with respect to the initial frailty $F_)$:
\begin{align*}
    &\frac{\partial}{\partial F_0} F(t;F_0) = \frac{N}{F_0 + (N-F_0)e^{-rNt}} - \frac{NF_0 \left(1- e^{-rNt}\right)  }{\left(F_0 + (N-F_0)e^{-rNt}\right)^2} >0, \\
    &\frac{\partial^2}{\partial F_0^2} F(t;F_0) = \frac{-2N\left(1-e^{-rNt}\right)}{\left(F_0 + (N-F_0)e^{-rNt}\right)^2} + \frac{2NF_0 \left(1- e^{-rNt}\right)^2  }{\left(F_0 + (N-F_0)e^{-rNt}\right)^3} <0.
\end{align*}
Thus, $F(t;F_0)$ is increasing and concave in $F_0$ for all $t>0$ and $0<F_0<N$. Since the force of mortality is proportional to the number of dysfunctional subsystems, it is also increasing and concave in $F_0$. Let $\overset{\circ}{e}_{x_0}(F_0)$ denote the complete expected future lifetime at age $x_0$ given initial frailty $F_0$:
\begin{align*}
 \overset{\circ}{e}_{x_0}(F_0) =   \mathbb{E}[T_{x_0}(F_0)|F_0] = \int_0^{\infty} \exp\left(-\int_0^t\mu_{x_0}(s;F_0)\diff s \right) \diff t.
\end{align*}
Differentiating under the integral sign yields
\begin{align*}
    &\frac{\partial}{\partial F_0} \overset{\circ}{e}_{x_0}(F_0) =  \int_0^{\infty} \left( -\int_0^t \left(\frac{\partial}{\partial F_0}\mu_{x_0}(s;F_0)\right)\diff s \right)\times \exp\left(-\int_0^t\mu_{x_0}(s;F_0)\diff s \right) \diff t < 0\\
    &\frac{\partial^2}{\partial F_0^2} \overset{\circ}{e}_{x_0}(F_0) = \int_0^{\infty} \left( -\int_0^t \left(\frac{\partial^2}{\partial F_0^2}\mu_{x_0}(s;F_0)\right)\diff s \right)\times \exp\left(-\int_0^t\mu_{x_0}(s;F_0)\diff s \right) \diff t\\
    & \qquad \qquad \qquad \qquad \qquad + \int_0^{\infty} \left( -\int_0^t \left(\frac{\partial}{\partial F_0}\mu_{x_0}(s;F_0)\right)\diff s \right)^2\times \exp\left(-\int_0^t\mu_{x_0}(s;F_0)\diff s \right) \diff t >0
\end{align*}
Hence, $ \overset{\circ}{e}_{x_0}(F_0)$ is decreasing and convex in $F_0$, and by Jensen's inequality, we have
\begin{align*}
    \mathbb{E}\left[ \overset{\circ}{e}_{x_0}(F_0)] \right] \geq \overset{\circ}{e}_{x_0}(\mathbb{E}[F_0]).
\end{align*}
This shows that heterogeneity in initial frailty increases average remaining lifetime compared with a homogeneous population having the same initial frailty.

\section{Results from male mortality data}\label{app_male}
\begin{figure}[ht!]
    \centering
    \includegraphics[width=0.75\linewidth]{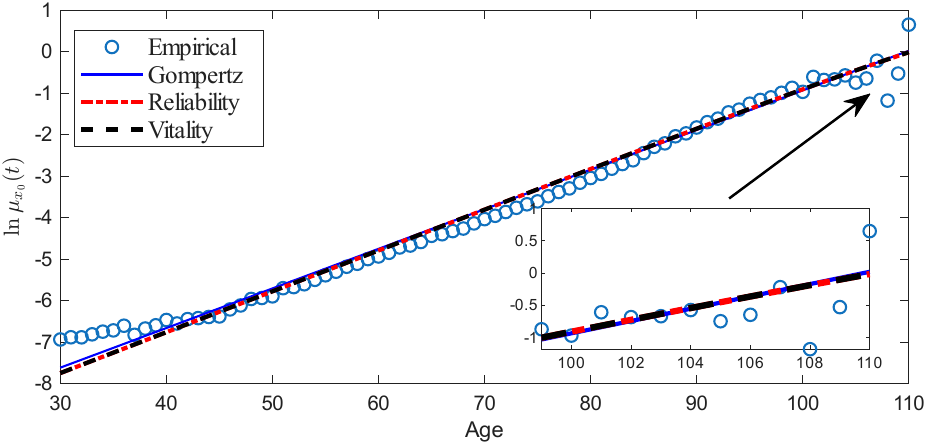}
    \caption{Observed mortality rates compared to the fitted Gompertz, reliability, and vitality models under the baseline specifications with male mortality data.}
    \label{fig:gompertz_benchmark_male}
\end{figure}

\begin{figure}[ht!]
    \centering
    \includegraphics[width=0.75\linewidth]{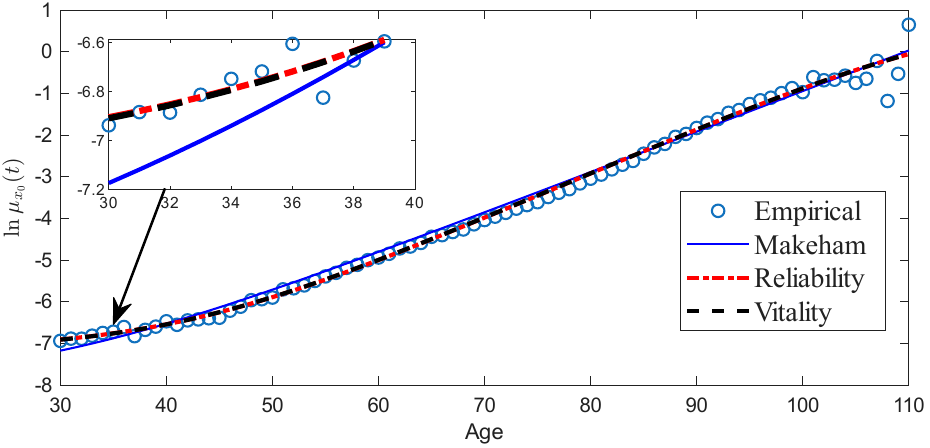}
    \caption{Fitted mortality curves under the Makeham Law and the extension from the reliability and vitality models with male mortality data.}
    \label{fig:gompertz_makeham_male}
\end{figure}

{For completeness, we repeat the baseline and Makeham calibrations of Section~\ref{sec:Analysis} using male mortality data. The estimated parameters are reported in Tables~\ref{tab:gompertz_parameters_male} and~\ref{tab:gompertz_makeham_male}, and the corresponding fitted curves are shown in Figures~\ref{fig:gompertz_benchmark_male} and~\ref{fig:gompertz_makeham_male}.}

\begin{table}[ht!]
    \centering
    \begin{tabular}{c|ccc}
    \toprule    
    Parameter   & Gompertz Law & Gompertz-Reliability Model & Gompertz-Vitality Model \\
    \midrule
        $b$        & $4.9257\times 10^{-4}$ & $4.3190\times 10^{-4}$ & $4.2612\times 10^{-4}$ \\
        $c$        & $1.1002$               & $1.1036$               & $1.1036$               \\
        $F_0$      & --                     & $62.0117$             & --                     \\
        $\alpha$   & --                     & --                     & $70.7367$             \\
    \midrule
        RSE        & $3.2252$               & $3.2017$               & $3.2017$           \\
    \bottomrule
    \end{tabular}
    \caption{Estimated parameters and relative squared error (RSE) from the fitted Gompertz law and Gompertz-specification of the reliability and vitality models, using male mortality data.}
    \label{tab:gompertz_parameters_male}
\end{table} 

\begin{table}[ht!]
    \centering
    \begin{tabular}{c|ccc}
    \toprule
    Parameter   & Makeham Law & Makeham-Reliability Model & Makeham-Vitality Model \\
         \midrule
      $\beta$   &  $3.4423\times 10^{-4}$ & $7.8456\times 10^{-4}$  & $7.4513\times 10^{-4}$ \\ 
      $b$       &  $4.2327\times 10^{-4}$ & $2.1913\times 10^{-4}$ & $2.1234\times 10^{-4}$ \\
      $c$       &  $1.1024$ & $1.1166$ & $1.1164$ \\
      $F_0$     &  -- & $83.8768$ & -- \\
      $\alpha$  &  -- & -- & $23.9516$ \\
      \midrule
      RSE       &  $3.1911$ & $3.0730$ & $3.0730$ \\
      \bottomrule
    \end{tabular}
    \caption{Estimated parameters and relative squared error (RSE) for the fitted Makeham Law and Makeham-specification of the reliability and vitality models, using male mortality data.}
    \label{tab:gompertz_makeham_male}
\end{table}

\end{appendices}

\section*{Acknowledgements}

Kenneth Q. Zhou acknowledges the support from the Natural Sciences and Engineering Research Council of Canada (RGPIN-2025-04157 and DGECR-2025-00488). Xiaobai Zhu acknowledges the support from the Research Grants Council of the Hong Kong Special Administrative Region, China (CUHK 24615523), and National Natural Science Foundation of China (No. 12301613).

\noindent

\end{document}